\documentclass[12pt]{JHEP3} 

\usepackage{amsmath,amssymb,axodraw}

\newcommand\fverb{\setbox\pippobox=\hbox\bgroup\verb}
\newcommand\fverbdo{\egroup\medskip\noindent%
                        \fbox{\unhbox\pippobox}\ }
\newcommand\fverbit{\egroup\item[\fbox{\unhbox\pippobox}]}
\newbox\pippobox

\title{Towards 5D Grand Unification without SUSY Flavor Problem}

\author{Ki-Young Choi${}^{a,b}$, Jihn E. Kim${}^{a,b}$
        and Hyun Min Lee${}^{a}$ \\ 
        ${}^a$ Physikalisches Institut der Universit$\ddot a$t Bonn, 
        Nussallee 12, D-53115 Bonn, Germany\\
        ${}^b$ School of Physics and Center for Theoretical Physics,
        Seoul National University, Seoul 151-747, Korea \\
        E-mail: \email{ckysky,jekim,minlee@th.physik.uni-bonn.de,
        jekim@phyp.snu.ac.kr}
}

\preprint{\hepph{0303213}}      

\abstract{We consider the renormalization group
approach to the SUSY flavor problem
in the supersymmetric $SU(5)$ model with one extra dimension.
In higher dimensional SUSY gauge theories, it has been recently
shown that power corrections due to 
Kaluza-Klein states of gauge fields run the soft masses generated
at the orbifold fixed point
to flavor conserving values in the infra-red limit.
In models with GUT breaking at the brane
where the GUT scale can
be larger than the compactification scale, 
we show that the addition of a bulk Higgs multiplet, 
which is necessary for the successful unification, is compatible 
with the flavor universality achieved at the compactification scale.
}


\begin{document} 


\section{Introduction}

Supersymmetry(SUSY) has been one of the most elegant candidates for solving 
the hierarchy problem in the Standard Model(SM)\cite{susy}. 
However, the general SUSY breaking with soft mass terms would give rise 
to over one hundred independent new parameters, 
which would lead to less predictivity.  
Moreover, there could appear flavor changing  
processes or $CP$ violating processes, which brings us to the   
so called the SUSY flavor problem\cite{masiero,masiero2}. 
In the general hidden sector model for SUSY breaking,  
the K$\ddot {\rm a}$hler potential in the supergravity Lagrangian contains 
counterterms of $O(1)$ coupling between visible and hidden sectors,
which give rise to flavor dependent scalar masses upon SUSY breaking. 
This can be understood from the fact that with flavor violation coming
from the Yukawa sector, the running squark mass has a logarithmic dependence
on the renormalization scale\cite{martin}. 
To avoid the SUSY flavor problem, there have been suggestions for the SUSY
breaking mechanism: gauge-mediation\cite{gaugem}, effective SUSY\cite{effsusy},
anomaly-mediation\cite{anom} and etc.  

For recent years, there have been alternatives to SUSY in the brane models 
with compact extra dimension(s) to explain the hierarchy problem 
without supersymmetry\cite{arkani,rs}. The common assumption 
in these approaches is that the gauge and matter fields are confined 
on the (3+1)-dimenisional subspace, so called 3-brane, 
embedded in $D>4$ dimensions, 
and the higher-dimensional fundamental scale or the physical scale 
at the brane can be of order of TeV scale 
without introducing a large hierarchy. 
However, it is difficult to maintain 
the idea of Grand Unification in these brane models with a low energy scale.

On the other hand, there were other attempts in the brane models to accommodate
the Grand Unification at the intermediate scale  
by considering the power running of gauge couplings due to the Kaluza-Klein(KK) 
modes of bulk gauge and/or matter fields between the GUT scale 
and the compactification scale\cite{DDG,ddg2}. 
In this case, supersymmetry should be assumed for naturalness of introducing 
the intermediate GUT scale. Therefore, we have to assume 
the higher dimensional SUSY for the Grand Unification in the brane models.

This idea of power running has been extended to the case with soft mass 
parameters in the higher dimensional GUT theories with broken 
SUSY\cite{kubo,cck}. 
It has been shown that for the bulk GUT theory with only gauge fields 
compactified on an orbifold, 
the soft masses at the orbifold fixed points 
powerly run to flavor conserving values in the infra-red limit 
going down to the GUT scale, due to the KK contributions of gauge fields  
to the renormalization group equations(RGEs). 
In higher than five dimensions, the consistent models in 6D are 
restricted to some of $SO(2N)$ models and exceptional groups due to the 
anomaly cancellation condition\cite{cck}. 
Moreover, in higher than six dimensions, 
the bulk supersymmetry becomes of ${\cal N}=4$ in 4D and then the ${\cal N}=4$ 
super Yang-Mills multiplet gives rise to a zero power-like beta function 
of the gauge coupling. 

In this paper, 
we reconsider the renormalization group approach to the SUSY flavor problem 
in 5D models on $S^1/Z_2$ where the GUT scale can be larger than the 
compactification scale. 
Then, we show that the non-universal KK contributions 
to the soft masses between the GUT
scale and the compactification scale are flavor diagonal. 
We can also consider the matter fields
in the bulk as far as the beta function coefficient of the zero-mode 
gauge coupling is power-like and negative. 
In fact, introduction of bulk matter fields is necessary for the more
successful gauge unification. 
Particularly, the case with the down type Higgs in the bulk 
and the up type Higgs on the 
brane was considered before for explaining the top-bottom mass 
hierarchy\cite{kkl}.
In this case, however, there appear additional power 
corrections to the RGEs due to the KK modes of the bulk Higgs 
which are flavor dependent. 
Nonetheless, we show that the additionally generated flavor violation is 
negligible thanks to the small down type Yukawa coupling. On the other hand,
we get a suppression of new $CP$ phases 
coming from the soft terms given at $M_{GUT}$ up to $10^{-2}$  
in either case without or with one bulk Higgs. 

The ${\cal N}=2$ SUSY in 5D is broken to ${\cal N}=1$ after orbifold 
compactification.
Then, we regard the final ${\cal N}=1$ SUSY and the GUT symmetry 
as being broken at the branes. 
In this case, the mass spectrum of the KK modes of bulk gauge 
fields are modified due to the brane mass terms\cite{ah,nomura,hdkim}. 
In particular, in the presence of the GUT breaking larger
than the compactification scale, the lowest $X,Y$ gauge bosons obtain masses
of order the compactification scale. When the GUT scale is of order the 
conventional 4D SUSY GUT scale, the lower $X,Y$ gauge boson masses could 
give rise to the rapid proton decay. 
However, since the wave functions of bulk $X,Y$ gauge bosons are suppressed 
at the brane with GUT breaking, the effective suppression scale 
of dimension-six proton decay operator becomes of order the GUT scale.  

Our paper is organized as follows. In the next section, we present our model
setup with localized SUSY and GUT breakings. Then, in section 3, we give 
the mass spectrums and wave functions of KK modes of the whole bulk
gauge multiplet and those of bulk Higgs multiplets in the presence of the 
localized symmetry breakings. In section 4, 
we consider the running of gauge couplings and comment on the proton decay 
problem in the models without or with bulk Higgs fields. 
In section 5, we go on to discuss the mass correction
of the GUT scalar on the brane due to the modified KK modes. 
In section 6, we also present the renormalization
group equations for the soft mass parameters above and below the GUT scale
and then show that the soft masses converge more 
into flavor conserving fixed points
in going down to the compactification scale.
In section 7, we present the RGEs 
for the case with bulk Higgs fields
and discuss the SUSY flavor problem in the model with one bulk Higgs field
in view of the new sources of flavor violation in RGEs. 
Finally, we will come to an end with the conclusion.

\section{Setup}

Let us consider a 5D SUSY SU(5) GUT where
one extra dimension 
is compactified on $S^1/Z_2$ with the radius $R$. 
It gives two fixed points at $y=0$ and
$y=\pi R$. We assume that in the bulk there exists only the 
${\cal N}=2$ $SU(5)$ gauge fields,
which contain an ${\cal N}=1$ vector multiplet $(A_\mu, \lambda_1)$
and an ${\cal N}=1$ chiral multiplet $(\Phi,\lambda_2)$.
All the other matter and Higgs multiplets are assumed to live at
the fixed point $y=0$.
The orbifold boundary condition breaks the
bulk ${\cal N}=2$ supersymmetry
down to the ${\cal N}=1$ supersymmetry, but
does not break the bulk gauge symmetry.

Including the ${\cal N}=2$\footnote{Let us borrow the notations of Mirabelli
and Peskin's\cite{peskin}} SUSY $SU(5)$ gauge fields in the bulk 
and the ${\cal N}=1$ SUSY SM matter fields($3({\bf {\bar 5}}+{\bf 10})$)
and Higgs fields\footnote{We will also consider the case with Higgs field(s)
propagating in the bulk. In this case, each Higgs multiplet comes 
as the zero mode of a bulk hypermultiplet.}
($H=\bf 5_H$, ${\bar H}=\bf {\bar 5}_H$) on the brane at $y=0$, 
the 5D action is given by 
\begin{eqnarray}
S_0&=&\int d^4x dy \bigg({\cal L}_5+\delta(y){\cal L}_4\bigg), \\
{\cal L}_5&=&{\rm Tr}\bigg[-\frac{1}{2}(F_{MN})^2 
+(D_M \Phi)^2+(\bar{\lambda}_i i\gamma^M D_M \lambda^i) 
+({\vec X})^2-\bar{ \lambda}_i[\Phi,\lambda^i]\bigg], \label{bulkaction} \\
{\cal L}_4&=&\int d^2\theta d^2{\bar\theta}
\bigg(\Psi_i({\bf 10})^\dagger e^{2g_5 V}\Psi_i({\bf 10})
+\Psi_i({\bf {\bar 5}})^\dagger e^{-2g_5 V}\Psi_i({\bf {\bar 5}}) \nonumber \\
&+&H^\dag e^{2g_5 V} H+{\bar H}^\dag e^{-2g_5 V} {\bar H}\bigg) \nonumber \\
&+&\bigg[\int d^2\theta \bigg(\frac{Y^{ij}_U}{4} H\Psi_i({\bf 10})
\Psi_j({\bf 10})
+\sqrt{2}Y^{ij}_D{\bar H}\Psi_i({\bf 10})\Psi_j({\bf {\bar 5}})\bigg)
+h.c.\bigg]\label{maction}
\end{eqnarray}
where we omitted the contraction of group indices, and 
$i,j$ run over 3 families of the Standard Model and $g_5$ is the 5D
$SU(5)$ gauge coupling with mass dimension $-\frac{1}{2}$. 
Note that in our convention, the covariant derivatives are defined as
$D_M \Phi=\partial_M-ig_5[A_M,\Phi]$(similarly for 
$\lambda^i$) 
and $D_\mu \Psi_i=(\partial_\mu-ig_5 A_\mu)\Psi_i$. The auxiliary field $D$ 
belonging to the ${\cal N}=1$ vetor multiplet on the boundary is given by
$D=X^3-\partial_y\Phi$. 

Now let us introduce 
a Higgs field ${\bf 24}=\Sigma$ at $y=0$ to break the GUT symmetry into the
SM one 
\begin{eqnarray}
{\cal L}_\Sigma=\delta(y)\bigg[\int d^2\theta d^2{\bar\theta}\,
2{\rm Tr}(\Sigma^\dag e^{2g_5 V}\Sigma e^{-2g_5 V})
+\bigg(\int d^2\theta\, W(\Sigma)+h.c.\bigg)\bigg]
\end{eqnarray}
with
\begin{eqnarray}
W(\Sigma)=\frac{1}{2}\mu_\Sigma{\rm Tr}
\Sigma^2+\frac{1}{3}Y_\lambda{\rm Tr}\Sigma^3
+Y_\Sigma H\Sigma\bar{H}+\mu_H H\bar{H}
\end{eqnarray}
where $Y_\lambda$ and $Y_\Sigma$ are dimensionless 
parameters, and $\mu_\Sigma$ and $\mu_H$ are dimensionful 
parameters\footnote{For the case with one(two) bulk Higgs multiplet(s), 
the dimensions of couplings become 
$[Y_{\Sigma}]=-\frac{1}{2}(-1)$ and $[\mu_H]=\frac{1}{2}(0)$.}.
Then, the $SU(5)$ GUT symmetry is broken by a vacuum expectation value of the
$\Sigma$ field 
\begin{equation}
\langle \Sigma \rangle = V \left( \begin{array}{ccccc}
                2&&&&\\
                &2&&&\\
                &&2&&\\
                &&&-3&\\
                &&&&-3  \end{array} \right)
\end{equation}
with $V=\mu_\Sigma/Y_\lambda$,
which gives brane masses to $X,Y$ gauge bosons 
\begin{eqnarray}
M_X=M_Y=(5\sqrt{2} g_5 V)^2\equiv M_V.
\end{eqnarray}
For $\mu_H=3Y_\lambda V$, the SM Higgs doublets are massless 
while the color-triplet Higgs fields are 
superheavy as 
\begin{eqnarray}
M_{H_C}=M_{H_{\bar C}}=5Y_\lambda V \label{chiggs} 
\end{eqnarray}
which will be denoted by
$\kappa$ later in the case with bulk Higgs multiplets. 
In the case with one(two) bulk Higgs multiplet(s), 
$\kappa$ has a mass dimension of $\frac{1}{2}(0)$. 

The SM group components of the adjoint Higgs multiplet
also obtain masses of the GUT scale after the GUT
breaking\footnote{$\Sigma^{(8,1)}$ and
$\Sigma^{(1,3)}$ have $M_\Sigma\equiv 5\mu_\Sigma/2$ and $\Sigma^{(1,1)}$
has mass $0.2M_\Sigma$.}.
On the other hand, for the broken components of the adjoint
Higgs, say the $X,Y$ directions, 
Goldstone bosons coming from $\Sigma^{({\bar 3},2)}$ and $\Sigma^{(3,2)}$ 
are eaten up by the $X,Y$ gauge bosons while
their physical components and the $X,Y$ gauginos with adjoint higgsino 
components get masses of $M_V$ 
to make up an ${\cal N}=1$ massive vector multiplets together 
with the zero modes of $X,Y$ gauge bosons. Since all the KK modes of bulk gauge 
fields are coupled to the adjoint Higgs fields on the brane, 
masses of KK ${\cal N}=1$ massive $X,Y$ gauge multiplets 
are also strongly affected by the GUT breaking effect on the brane. 

Next let us add soft SUSY breaking terms for the brane fields 
at $y=0$ as in the 4D SUSY SU(5) case except the gaugino mass terms. 
We assume that the gauge invariant soft masses 
for gauginos generically
appear at both fixed points with arbitrarily different values as
\begin{eqnarray}
{\cal L}_{soft}\supset -\delta(y)(\frac{1}{2}\varepsilon_0\lambda_1\lambda_1
+h.c.) 
-\delta(y-\pi R)(\frac{1}{2}\varepsilon_\pi\lambda_1\lambda_1+h.c.)
\end{eqnarray} 
where $\varepsilon_{0,\pi}$ are dimensionless quantities 
of ${\cal O}(10^{-12})$ to represent the weak scale supersymmetry breaking.
Note that we do not include soft terms for $\lambda_2$ at the fixed points
because those will not affect the equations of motion for gauginos due
to the odd parity of $\lambda_2$ on $S^1/Z_2$. 
$\lambda_2$ only participates in modifying the mass
spectrum via mixing with $\lambda_1$ in the bulk. For softness as will be shown 
later, we assume that $\varepsilon_0$ is zero. 
In the case with bulk Higgs fields, we introduce soft mass terms for 
bulk Higgs scalars only at $y=\pi R$ for the same reason.

\section{Mass spectrum with brane-induced SUSY and gauge symmetry breakings}

Firstly, let us give a brief review on
the KK mass spectrum of the $X,Y$ gauge bosons in the presence 
of their localized gauge symmetry breaking\cite{nomura,hdkim}.
After the GUT breaking by the VEV of $\Sigma$,
the $X,Y$ gauge bosons acquire brane-localized masses,
\begin{eqnarray}
\mathcal{L} \supset \delta(y) \frac{1}{2}M_V A^{\hat a}_\mu(x,y) 
A^{{\hat a}\mu}(x,y)
\end{eqnarray}
where $\hat a$ runs over broken generators.

Then, under the KK reduction, the $X,Y$ gauge bosons for $-\pi R<y<\pi R$ 
can be written as 
\begin{eqnarray}
A^{\hat a}_\mu(x,y)=\frac{1}{\sqrt{\pi R}}\sum_n N_n A^{{\hat a}(n)}_\mu(x)
\cos(M^A_n |y|-\theta^A_n)\label{bosonwf}
\end{eqnarray} 
satisfying the equation of motion 
\begin{eqnarray}
-\partial^2_y A^{\hat a}_\mu(x,y)+\delta(y)
M_V A^{\hat a}_\mu(x,y)=(M^A_n)^2A^{\hat a}_\mu(x,y).
\end{eqnarray}
Consequently, with $\theta^A_n=M^A_n\pi R$, the mass spectrum is determined
by the boundary condition at $y=0$  
\begin{eqnarray}
\tan(M^A_n \pi R)=\frac{M_V}{2M^A_n}\label{gaugeevalue}.
\end{eqnarray}
In the limit $M_V\gg M^A_n$,
we can obtain the approximate masses as
\begin{eqnarray}
M^A_n\simeq\left(n+\frac12\right)M_c
\bigg\{1-\frac{2M_c}{\pi M_V }
+\bigg(\frac{2M_c}{\pi M_V  }\bigg)^2\bigg\}
\label{gaugemass}
\end{eqnarray}
where $M_c=1/R$ and $n=0,1,2,\cdots$. Therefore, even with the large gauge
boson masses on the brane, the lowest KK modes of $X,Y$ gauge bosons 
get masses of order the compactification scale. This fact is related 
to the modified wave function of
$X,Y$ gauge bosons, which are repelled from the brane at $y=0$ 
due to the brane mass term. Thus, it gives rise to the suppression of effective
gauge coupling at the brane, so that the $X,Y$ gauge boson mass 
of order $M_c$ much lower than the GUT scale can be consistent with the proton
stability\cite{nomura}.
When we integrate the 5D action for the gauge bosons over the extra dimension 
with eq.~(\ref{bosonwf}), the normalization constant $N_n$ is determined by
\begin{eqnarray}
N_n=\bigg(1+\frac{M_c M_V}{2\pi (M^A_n)^2}\cos^2\theta^A_n\bigg)^{-1/2}.
\end{eqnarray}
Since the SM gauge bosons do not have brane masses, their KK mass spectrum
is just given as $M_n=n M_c$ with $n=0,1,2,\cdots$ 
and the normalization becomes $N_n=\frac{1}{\sqrt {2^{\delta_{n,0}}}}$.

Let us take into account the mass spectrum of $X,Y$ gauginos in the presence of
their Dirac mass terms with $X,Y$ components of brane higgsinos at $y=0$ and 
their soft mass terms at both fixed points.
The mass eigenstates of $X,Y$ components of adjoint fermions 
will be written as a linear combination 
of the KK modes of $\lambda_{1,2}$ and $\tilde\Sigma$ 
due to bulk and brane mixings. 
To obtain the mass spectrum and the mass eigenstates of gauginos, 
we have to solve the following field equations
\begin{eqnarray}
&&i{\overline \sigma}^\mu\partial_\mu{\tilde\Sigma}^b
-\sqrt{M_V}\overline {\lambda^a_1}|_{y=0}=0 \\
&&i{\overline \sigma}^\mu\partial_\mu\lambda_2^a
+\partial_y\overline {\lambda^a_1}=0 \\
&&i{\overline \sigma}^\mu\partial_\mu\lambda^a_1
-\partial_y\overline {\lambda^a_2}
-(\sqrt{M_V}\overline{ {\tilde\Sigma}^b}+\varepsilon_0
\overline{\lambda^a_1})\delta(y)
-\varepsilon_\pi \overline{\lambda^a_1}\delta(y-\pi R)=0
\end{eqnarray}
where $a,b$ run over inequivalent pairs of two nearest indices of broken 
generators. 

Let us take the form of the gaugino solutions to be
consistent with the orbifold symmetry for $-\pi R<y<\pi R$,
\begin{eqnarray}
\lambda^a_1(x,y)&=&\frac{1}{\sqrt{2\pi R}}\sum_n K_n\eta_1^{a(n)}(x)
\cos(M^X_n |y|-\theta^X_n) \\
\lambda^a_2(x,y)&=&\frac{1}{\sqrt{2\pi R}}\sum_n K_n\eta_2^{a(n)}(x)
\epsilon(y)\sin(M^X_n |y|-\theta^X_n)
\end{eqnarray}
where $\epsilon(y)$ is a sign function.
If we also assume
\begin{eqnarray}
\eta^{a(n)}_1(x)&=&\eta^{a(n)}_2(x)\equiv \lambda^{a(n)}(x), \\
{\tilde \Sigma}^a(x)&=&\frac{1}{\sqrt{2\pi R}}\sum_n K_n 
\frac{\sqrt{M_V}}{M^X_n}\cos\theta^X_n
\lambda^{a(n)}(x),
\end{eqnarray}
then we obtain the solutions for $\theta^X_n$ and $M^X_n$ from the boundary 
conditions at the branes 
\begin{eqnarray}
\tan\theta^X_n&=&\frac{M_V}{2M^X_n}+\frac{\varepsilon_0}{2}, 
\label{gauginotheta}\\
\tan(M^X_n \pi R-\theta^X_n)&=&\frac{\varepsilon_\pi}{2}.
\end{eqnarray}
Therefore, the eigenvalue equation for the $X,Y$ gaugino masses are 
\begin{eqnarray}
\tan(M^X_n\pi R-{\rm arctan}(\frac{\varepsilon_\pi}{2}))
=\frac{M_V}{2M^X_n}+\frac{\varepsilon_0}{2} \label{gauginomass}.
\end{eqnarray}
Thus, we obtain the approximate mass spectrum of $X,Y$ gauginos as
\begin{eqnarray}
M^X_n&\simeq& M^{X(0)}_n\bigg\{1-\frac{2M_c}{\pi M_V}
\bigg(1+\frac{\varepsilon_0 M^{X(0)}_n}{M_V}\bigg)^{-1} \nonumber \\
&+&\bigg(\frac{2M_c}{\pi M_V}\bigg)^2 
\bigg(1+\frac{\varepsilon_0 M^{X(0)}_n}{M_V}\bigg)^{-3}\bigg\}
\end{eqnarray}
where
\begin{eqnarray}
M^{X(0)}_n=M_c\bigg[(n+\frac{1}{2})+\frac{1}{\pi}{\rm arctan}
(\frac{\varepsilon_\pi}{2})\bigg]
\end{eqnarray}
with integer $n$.
We can also write the 4D effective action for the adjoint fermions 
in terms of the mass eigenstates $\lambda^{a(n)}(x)$ to obtain the normalization
constant as
\begin{eqnarray}
K_n=\bigg(1+\frac{M_c M_V}{2\pi (M^X_n)^2}
\cos^2\theta^X_n\bigg)^{-1/2}\label{gauginonor}.
\end{eqnarray} 
The mass spectrum of the SM gauginos can be obtained by nullifying the GUT 
breaking on the brane in eqs.~(\ref{gauginotheta}), (\ref{gauginomass})
and (\ref{gauginonor}) as  
\begin{eqnarray}
\tan\theta^{SM}_n&=&\frac{\varepsilon_0}{2}, \\
\tan(M^{SM}_n\pi R-{\rm arctan}(\frac{\varepsilon_\pi}{2}))
&=&\frac{\varepsilon_0}{2}, \\
K^{SM}_n&=&1.
\end{eqnarray}
Then, we obtain the SM gaugino mass spectrum as
\begin{eqnarray}
M^{SM}_n&=&n M_c+\frac{M_c}{\pi}{\rm arctan}(\frac{\varepsilon_0}{2})
+\frac{M_c}{\pi}{\rm arctan}(\frac{\varepsilon_\pi}{2}) \nonumber \\
&=&n M_c+\frac{M_c}{\pi}{\rm arctan}
\bigg[\frac{\varepsilon_0+\varepsilon_\pi}
{2-\frac{\varepsilon_0\varepsilon_\pi}{2}}\bigg]
\end{eqnarray} 
where $n$ is an integer. 

Let us remark on the restoration of supersymmetry in our setup.
When we take the SM gaugino masses with opposite signs
at the branes, $\varepsilon_0=-\varepsilon_\pi$, there is no modification
of the KK masses, which implies that supersymmetry can be restored for the SM
gauginos
even if the KK modes are mixed to make up the mass eigenstates. However, this
is not the case for the $X,Y$ gauginos. Even with 
$\varepsilon_0=-\varepsilon_\pi$, there is generically no restoration of
supersymmetry for the $X,Y$ gauginos 
since there is the mixing between gauginos and Higgsinos at $y=0$. 

Furthermore we comment on the couplings of gauge bosons and gauginos in view
of the modified wave functions. To begin with, let us take into account the
case with the SM gauge fields. In the presence of nonzero gaugino masses 
at both fixed points, the coupling of the KK mode of SM gauginos 
to brane matters at $y=0$ is proportional to $\cos\theta^{SM}_n$, which is
generically 
different from that of the corresponding KK mode of the SM gauge 
bosons. Then, softness of the SM gaugino masses on the brane can be 
guaranteed only if $\cos\theta^{SM}_n=1$, i.e., there is no gaugino mass 
at $y=0$.
Therefore, for our discussion, henceforth we assume 
$\varepsilon_0=0$. On the other hand, for each pair of  
$X,Y$ gauge bosons and gauginos, we also have different couplings 
at $y=0$ proportional
to $N_n \cos\theta^A_n$ and $K_n\cos\theta^X_n$, respectively. 
Even with $\varepsilon_0=0$, however, 
couplings of $X,Y$ gauge bosons and gauginos 
are not the same, so there would appear quadratic divergences to soft masses 
for brane fields from loop corrections. 
Let us postpone the detailed discussion on this issue to
the section 5. 

For completeness with gauge multiplet, 
we also add the mass spectrum for the bulk real 
scalar field $\Phi$ coupled to the adjoint Higgs $\Sigma$ on the brane. 
The relevant Lagrangian is
\begin{eqnarray}
\mathcal{L} \supset&& \frac12 D^aD^a+D^a(\partial_y \Phi^a) 
+\frac12(\partial_\mu \Phi^a)(\partial^\mu\Phi^a) \nonumber \\
&+&\delta(y)[(\partial_\mu{\rm Im}\Sigma^a)^2+\sqrt{2M_V}D^b{\rm Im}\Sigma^a]
\end{eqnarray}
where the last term is nonzero only for $(a,b)$ of two nearest indices 
of broken generators
and we wrote the action in terms of the auxiliary
field on the boundary, $D^a=X^{3a}-\partial_y \Phi^a$.
Here we note that the real part of the $X,Y$ components of $\Sigma$ 
can be gauged away 
to give the longitudinal degree of freedom to the $X,Y$ gauge bosons 
while the imaginary part of the $X,Y$ components of $\Sigma$ gets mass
due to the mixing with $\Phi$ at the brane. 
The equations of motion for the $X,Y$ components of scalars become 
\begin{eqnarray}
&&D^a+\partial_y\Phi^a+\sqrt{2M_V}{\rm Im}\Sigma^b\delta(y)=0 \\
&& -\partial^\mu\partial_\mu\Phi^a-\partial_yD^a=0\\
&& -2\partial^\mu\partial_\mu{\rm Im}\Sigma^b+\sqrt{2M_V}D^a|_{y=0}=0.
\end{eqnarray}
Then, we can find the solutions for the scalars 
for $-\pi R < y< \pi R$ as
\begin{eqnarray}
\Phi^a(x,y)&=&\frac{1}{\sqrt{\pi R}}\sum_n C_n
\epsilon(y)\sin(M^\Phi_n |y|-\theta^\Phi_n)\varphi^{a(n)}(x),\\
{\rm Im}\Sigma^b(x)&=&\frac{1}{\sqrt{\pi R}}\sum_n C_n
\frac{\sqrt{M_V}}{M^\Phi_n}(\cos\theta^\Phi_n)\varphi^{a(n)}(x), \\ 
D^a(x,y)&=&\frac{1}{\sqrt{\pi R}}\sum_nC_n
\cos(M^\Phi_n|y|-\theta^\Phi_n)D^a_n(x),
\end{eqnarray}
where $\theta^\Phi_n=\theta^A_n=M^A_n\pi R$ 
and $M^\Phi_n=M^A_n$ from the boundary condition at the brane 
and the normalization 
constant $C_n$ is also the same as $N_n$ for the $X,Y$ gauge bosons.
Therefore, the mass spectrum for the bulk real 
scalar field $\Phi$ is the same as the one for the gauge bosons 
as required from the 4D ${\cal N}=1$ massive supersymmetry. 

From the action eq.~(\ref{maction}), 
the matter scalar $\phi$ also couples to the auxiliary field $D$. 
Then, we can integrate out the KK modes of the auxiliary field $D$ by using
the following equation of motion 
\begin{eqnarray}
D^a_n+M^A_n\varphi^a_n+\frac{g_5}{\sqrt{\pi R}}N_n(\cos\theta^A_n)
\phi^\dag T^a\phi=0.
\end{eqnarray} 
This gives the interaction term between the matter scalar $\phi$ and
the KK modes of the bulk real scalar $\varphi^{(n)}$ and the self interaction
term as
\begin{eqnarray}
\int d^4 x\sum_n\bigg[-\frac{g_5}{\sqrt{\pi R}}N_nM^A_n
(\cos\theta^A_n)\phi^\dag \varphi^{(n)}\phi
-\frac{g^2_5}{2\pi R}N_n^2(\cos\theta^A_n)^2(\phi^\dag T^a\phi)^2\bigg].
\end{eqnarray}

Now let us close this section with the case that Higgs multiplets 
propagate in the bulk. The KK mass spectrum of
colored Higgs triplets is also modified due to the brane mass term 
after the GUT breaking, which can be written in terms of 
$4D$ ${\cal N}=1$ superfields as $(\kappa H_C H_{\bar C}+h.c)\delta(y)$
where $\kappa\simeq 5Y_\Sigma V$ from eq.~(\ref{chiggs}) in case of 
$\mu_H=3Y_\lambda V$ being not exact. 
On the other hand, the KK mass
spectrum of the bulk Higgs doublets is also modified for the nonvanishing brane
mass term, $(\varepsilon_H H_u H_d+h.c.)\delta(y)$ where 
$\varepsilon_H=\mu_H-3Y_\lambda V$.

The case with two Higgs 
multiplets($H=H_C+H_u, {\bar H}=H_{\bar C}+H_d$) in the bulk, 
for which $\kappa$ and $\epsilon_H$ are dimensionless,
was dealt with in Ref.~\cite{nomura}. In the presence of the brane mass term
for the bulk colored higgsinos, it has been shown 
that the zero modes of two bulk higgsinos get a Dirac mass. 
From the eigenvalue equation, $\tan^2(M^{H_C}_n\pi R)=\kappa^2/4$ 
given in \cite{nomura},
the KK mass spectrum of colored higgsinos is given by 
\begin{eqnarray}
M^{H_C}_n=nM_c+\frac{M_c}{\pi}{\rm arctan}\bigg(\frac{\kappa}{2}\bigg)
\end{eqnarray}
where $n$ is an integer. Likewise, the KK mass spectrum of the higgsino 
doublets is the one with $\kappa$ replaced by $\varepsilon_H$ in the above 
equation. Then, the lowest higgsino doublet mass is approximately given by 
$\varepsilon_H M_c/(2\pi)$, which corresponds to the $\mu$ parameter 
of order TeV in the MSSM after a fine-tuning between GUT parameters.

Here we also consider the case with one Higgs($H$) on the brane 
and the other Higgs($\bar H$) in the bulk, in which case $\kappa$ and 
$\varepsilon_H$ have
a mass dimension $\frac{1}{2}$. Without a brane mass term for higgsinos, 
$\bar H$ comes from the zero
mode of a bulk hypermultiplet composed of 
(${\bar H}=H_{\bar C}+H_d,{\bar H}^c=H^c_{\bar C}+H^c_d$).
When we take into account the brane mass term for the colored higgsinos 
at $y=0$, which mixes brane and bulk colored higgsinos,  
the equations of motion for the brane and bulk colored higgsinos 
are given by
\begin{eqnarray}
& &i{\bar\sigma}^\mu\partial_\mu H_C-\kappa {\overline H}_{\bar C}|_{y=0}=0, \\
& &i{\bar\sigma}^\mu\partial_\mu H_{\bar C}-\partial_y {\overline H}^c_{\bar C}
-\kappa{\overline H}_C\delta(y)=0, \\
& &i{\bar\sigma}^\mu\partial_\mu H^c_{\bar C}
+\partial_y {\overline H}_{\bar C}=0.
\end{eqnarray}
Following the similar procedure as for the gauge multiplet, we find the 
solutions of the colored higgsinos for $-\pi R<y<\pi R$ as
\begin{eqnarray}
H_{\bar C}(x,y)&=&\frac{1}{\sqrt{\pi R}}\sum_n N^H_n h^{(n)}_1(x)
{\rm cos}(M^{H_C}_n (|y|-\pi R)),
\\
H^c_{\bar C}(x,y)&=&\frac{1}{\sqrt{\pi R}}\sum_n N^H_n h^{(n)}_2(x) 
\epsilon(y)\sin (M^{H_C}_n (|y|-\pi R)),
\\
H_C(x)&=&\frac{1}{\sqrt{\pi R}}\sum_n N^H_n \frac{\kappa}{M^{H_C}_n}
{\rm cos}(M^{H_C}_n\pi R)h^{(n)}_2(x),
\end{eqnarray} 
where 
$i{\bar\sigma}^\mu\partial_\mu h^{(n)}_1=M^{H_C}_n {\overline h}^{(n)}_2$, 
$i{\bar\sigma}^\mu\partial_\mu h^{(n)}_2=M^{H_C}_n {\overline h}^{(n)}_1$, 
and the normalization
constant $N_n^H$ is given by
\begin{eqnarray}
N^H_n=\bigg(1+\frac{M_c\kappa^2}{2\pi(M^{H_C}_n)^2}
\cos^2(M^{H_C}_n\pi R)\bigg)^{-1/2}.
\end{eqnarray}
Accordingly we also obtain the eigenvalue equation as
\begin{eqnarray}
{\rm tan}(M^{H_C}_n\pi R)=\frac{\kappa^2}{2M^{H_C}_n}.\label{higgseval}
\end{eqnarray}
Here we note that a pair of Weyl spinors, $h^{(n)}_1$ and $h^{(n)}_2$, 
make up a Dirac mass at each KK level. The brane higgsino participates in
the mixing between KK modes to make the zero mode of the bulk higgsino massive. 
Therefore, for $\kappa^2\gg M^{H_C}_n$, the approximate mass spectrum for the 
colored higgsinos is given by
\begin{eqnarray}
M^{H_C}_n\simeq \bigg(n+\frac{1}{2}\bigg)M_c\bigg\{1-\frac{2M_c}{\pi\kappa^2}
+\bigg(\frac{2M_c}{\pi\kappa^2}\bigg)^2\bigg\} 
\end{eqnarray}
where $n=0,1,2,\cdots$. This result reminds us of the case with gauge 
multiplet without SUSY breaking when we identify $\kappa^2$ as $M_V$.
Likewise, the eigenvalue equation for higgsino doublets is given by the one
with $\kappa$ replaced by $\varepsilon_H$ in eq.~(\ref{higgseval}). Then, 
for $\varepsilon^2_H\ll M^H_n$, the mass spectrum of higgsino doublets become
$M^H_0\simeq \varepsilon_H\sqrt{M_c/(2\pi)}$, and 
$M^H_n\simeq nM_c+\varepsilon^2_H/(2n\pi)$ for $n=1,2,3,\cdots$. 
We also need a fine-tuning between the GUT parameters to obtain the zero mode
higgsino mass around TeV.

The bulk Higgs scalars 
have the same modification due to the brane mass term at $y=0$ 
as that of bulk higgsinos due to the remaining ${\cal N}=1$ supersymmetry. 
Then, only taking into account the allowed gauge symmetry,  
we can put the SUSY breaking mass terms for the bulk Higgs scalar 
at both fixed points, which then make a shift of the KK spectrum 
of the bulk Higgs scalar. 
However, the SUSY breaking term at $y=0$ gives rise to different couplings 
of bulk Higgs and higgsino to a brane scalar as in the case with 
the bulk gauge multiplet, so that the quadratic divergences are not cancelled
between fermionic and bosonic loops at the brane. 
Therefore, we have to assume that the soft 
mass term for a bulk Higgs scalar appears only at $y=\pi R$.

\section{Gauge coupling unification and proton decay}

We have shown that the mass spectrum of
the bulk gauge multiplet is modified
in the existence of the mass terms coming from
the supersymmetry and gauge symmetry breakings on the brane.
In particular, even with the large scale masses
due to the GUT breaking at the brane, the ${\cal N}=1$
massive $X,Y$ gauge multiplet
get masses of order the compactification scale.
The brane soft mass term for the bulk
gauginos just gives rise to an overall small shift
for the KK spectrum of gauginos as in Figs.~ 1 and 2.

In this section, in our model  
that the compactification scale $M_c$ is smaller
than the GUT scale $M_{GUT}$, we consider the running of gauge
couplings due to the KK modes of the bulk gauge multiplet 
up to the GUT scale. In this case, however, 
we must guarantee the successful unification which is spoiled due to
the additional non-universal log running,  
as well as the proton longevity
which is challenged due to the somewhat smaller $X,Y$ gauge boson masses.

When we take into account the KK contributions
above the compactification
scale($M_c\equiv 1/R$) ignoring the
SUSY breaking effects on the KK spectrum,
the running of the zero-mode gauge couplings is given by
\begin{eqnarray}
\alpha^{-1}_i(\mu)&=&\alpha^{-1}_i(M_{GUT})+\frac{1}{2\pi}b'_i\ln
\bigg(\frac{M_{GUT}}{\mu}\bigg) \nonumber \\
&+&\frac{1}{2\pi}b^{SM}_i
\sum_{0<nM_c<M_{GUT}}\ln\bigg(\frac{M_{GUT}}{nM_c}\bigg)
+\frac{1}{2\pi} b^X_i\sum_{M_n<M_{GUT}}
\ln\bigg(\frac{M_{GUT}}{M_n}\bigg)\label{running}
\end{eqnarray}
where $M_n$ are the KK masses of the ${\cal N}=1$ massive
$X,Y$ gauge multiplet,
and $b'_i=(33/5,1,-3)$, $b^{SM}_i=(0,-4,-6)$, $b^X_i=(-10,-6,-4)$
are the beta function coefficients for the
zero mode MSSM fields, the ${\cal N}=1$ massive MSSM gauge multiplet
and the ${\cal N}=1$
massive X,Y gauge multiplet, respectively. Then, with $N=M_{GUT}/M_c$,
using the Stirling's formular, we can get the sum of the KK modes as
\begin{eqnarray}
\sum_{0<nM_c<M_{GUT}}\ln\bigg(\frac{M_{GUT}}{nM_c}\bigg)&=&\sum_{n=1}^{N-1}
\ln\bigg(\frac{N}{n}\bigg)\nonumber \\
&\simeq& N-\frac{1}{2}\ln(2\pi N), \\
\sum_{M_n<M_{GUT}}\ln\bigg(\frac{M_{GUT}}{M_n}\bigg)
&=&\sum_{n=1}^N\ln\bigg[\frac{2N}{(2n-1)(1-\zeta/N+(\zeta/N)^2)}\bigg]
\nonumber \\
&\simeq&N-\frac{1}{2}\ln 2+\zeta+{\cal O}(\frac{\zeta^2}{N})
\end{eqnarray}
where
\begin{eqnarray}
\zeta=\frac{2M_{GUT}}{\pi M_V}.
\end{eqnarray}
Therefore, the resulting running equations of low energy gauge couplings
are given by
\begin{eqnarray}
\alpha^{-1}_i(\mu)&=&\alpha^{-1}_i(M_{GUT})+\frac{1}{2\pi}b'_i\ln
\bigg(\frac{M_{GUT}}{\mu}\bigg)+\frac{1}{2\pi}bN \nonumber \\
&-&\frac{1}{4\pi}b^{SM}_i \ln(2\pi N)-\frac{1}{4\pi}b^X_i\ln 2
+\frac{1}{2\pi}b^X_i\zeta
\end{eqnarray}
where $b\equiv b^{SM}_i+b^X_i=-10$. The power
running proportional to $N$ is universal for all gauge
couplings and cannot contribute in narrowing down
the separations of the gauge couplings, but
it leads the gauge couplings to very small ones
at the unification scale since $b<0$. That is,
the zero-mode gauge coupling at the unification scale becomes
$\alpha(M_{GUT})\simeq -2\pi/(bN)\sim 6.3\times 10^{-3}$ 
for $N=10^2$. In our paper, we assume that the maximum 
number of KK modes($N$) is of ${\cal O}(100)$ for the validity 
of perturbative calculations. 
On the other hand, the log terms still contribute in narrowing down
the separations of gauge couplings toward a unified gauge coupling
constant at $M_{GUT}$.

A linear combination of the above running equations gives a
theoretic value of $\alpha_3$ as
\begin{eqnarray}
\alpha^{-1}_3=\frac{12}{7}\alpha^{-1}_2-\frac{5}{7}\alpha^{-1}_1
+\frac{\tilde b}{2\pi}\ln(\pi N)+\frac{\tilde c}{2\pi}\zeta \label{strong}
\end{eqnarray}
where ${\tilde b}=-\frac{1}{2}(b^{SM}_3-(12/7)b^{SM}_2+(5/7)b^{SM}_1)=-3/7$
and ${\tilde c}=b^X_3-(12/7)b^X_2+(5/7)b^X_1=-6/7$.
Then, we obtain the KK correction to the value of $\alpha_s$ with $N=10^2$
in the 4D minimal SUSY GUT as
\begin{eqnarray}\label{deltaKK}
\delta_{KK}\alpha_s\simeq-\frac{1}{2\pi}\alpha^2_s(M_Z)
({\tilde b}\ln(\pi N)+{\tilde c}\zeta)\simeq 0.0066
\end{eqnarray}
which gives the strong coupling at $M_Z$ in our 5D model 
as $\alpha_s^{KK}=\alpha^{SGUT,0}_s+\delta_{KK}\alpha_s$ 
where $\alpha^{SGUT,0}_s$ is the value from the 4D SUSY GUT without threshold
corrections.
Note, however, that this correction is not in
the favorable direction. Thus, we find that there is a competition 
between the large separation of scales $N$ and the correct value of $\alpha_s$. 
Currently, we have the experimental
data which is somewhat smaller than the MSSM prediction,
$\alpha^{exp}_s(M_Z)-\alpha^{SGUT,0}_s=-0.013\pm 0.0045$\cite{kkl}. 
Our KK modes positively add to the MSSM prediction as given in
(\ref{deltaKK}), and the discrepancy is little bit enlarged.
However, considering the current experimental error
bounds and the theoretically unknown threshold
corrections at the GUT scale, we can tolerate the positive
contribution of order 0.006 to the theoretical value
of $\alpha_s$. 

If we had put some of brane fields in the bulk, we could have
been in a better situation for the successful prediction 
for the strong coupling. For instance, let us consider the case (1)  
where the down-type Higgs multiplet comes from the bulk
and the up-type Higgs multiplet is put on the brane\cite{kkl},
and the case (2) where
two Higgs multiplets $5_H$ and ${\bar 5}_H$ come from the bulk. 
Then, as shown in the previous section, the mass 
spectrum of colored Higgs triplet(s) is modified due to the brane mass term. 
In the case (2), 
with $\kappa$ larger than $O(1)$, the approximate KK masses for the colored 
Higgs triplets are $M^{H_C}_n\simeq (n+1/2)M_c-2M_c/(\pi\kappa)$, $n\geq 0$. 
On the other hand, in the case (1), the approximate mass spectrum of the
colored Higgs triplet for $\kappa^2\gg M^H_n$ is 
$M^{H_C}_n\simeq (n+1/2)M_c(1-\zeta'/N+(\zeta'/N)^2)$ where 
$\zeta'=2M_{GUT}/(\pi \kappa^2)$.
But in both cases, the bulk Higgs doublet has the same mass spectrum 
as for the SM gauge multiplet. Therefore, there exist additional contributions 
coming from the KK modes 
of bulk Higgs multiplets on the right-hand side of eq.~(\ref{running}):
\begin{eqnarray}
\frac{1}{2\pi}b^{H_d}_i\sum_{0<nM_c<M_{GUT}}\ln\bigg(\frac{M_{GUT}}{M_n}\bigg)
+\frac{1}{2\pi}b^{H_C}_i\sum_{M^{H_C}_n<M_{GUT}}
\ln\bigg(\frac{M_{GUT}}{M^{H_C}_n}\bigg)
\end{eqnarray}   
where $b^{H_d}_i,b^{H_C}_i$ are beta function coefficients 
for the Higgs doublet and triplet, $b^{H_d}_i=(3/5(6/5),1(2),0)$, 
$b^{H_C}_i=(2/5(4/5),0,1(2))$ for one(two) bulk Higgs multiplet(s). 
Consequently, the additional KK contributions sum up to make the power
running a bit smaller, $b\rightarrow b+b^{H_d}_i+b^{H_C}_i=-9(-8)$. 
On the other hand, we find that the bulk Higgs correction 
together with the bulk gauge contribution is now in the
favorable direction toward the experimental value of the strong coupling. 
That is, in eq.~(\ref{strong}),
$\tilde b$ becomes ${\tilde b}+{\tilde b}^H=3/14(6/7)$ 
where 
${\tilde b}^H=-\frac{1}{2}(b^{H_d}_3-(12/7)b^{H_d}_2+(5/7)b^{H_d}_1)=9/14(9/7)$
for one(two) bulk Higgs multiplet(s), and $\tilde c\zeta$ 
becomes changed to ${\tilde c}\zeta+{\tilde c}_H\zeta'$ 
only for the case with one bulk Higgs
where ${\tilde c}_H=b^{H_C}_3-(12/7)b^{H_C}_2+(5/7)b^{H_C}_1=9/7$. 
Then, we obtain the KK correction
to the strong coupling as 
$\delta_{KK}\alpha_s\simeq -0.0015(-0.0095)$ 
for $N=10^2$. 
 
Now we can determine the unification scale in the minimal case with only gauge
fields in the bulk 
by using running equations for $\alpha_1$ and $\alpha_2$ as
\begin{eqnarray}
M_{GUT}&=&M_X (2\pi N)^{\frac{1}{2}\frac{b^{SM}_1-b^{SM}_2}{b'_1-b'_2}}
\,e^{\frac{b^X_1-b^X_2}{b'_1-b'_2}(\frac{1}{2}\ln2-\zeta)} \nonumber \\
&\simeq& 12 M_X\simeq 2.5 \times 10^{17} {\rm GeV}
\end{eqnarray}
where $M_X=M_Z \, e^{\frac{2\pi}{b'_1-b'_2}(\alpha^{-1}_1-\alpha^{-1}_2)}\simeq 2\times 10^{16}$ GeV is the unification scale in the MSSM and we took 
$\zeta=2/\pi$ for $M_V=M_{GUT}$. Therefore, the GUT scale
in our case is a bit higher than in the usual 4D SUSY GUTs. 
Then, with $N=M_{GUT}/M_c=10^2$, the compactification scale is given by 
$M_c\simeq 2.5\times 10^{15}$ GeV. 
On the other hand, in the case with one bulk Higgs,
similarly we can get the unification scale and the compactification scale 
for $N=10^2$ as $M_{GUT}\simeq 1.3\times 10^{17}$ GeV and 
$M_c\simeq 1.3\times 10^{15}$ GeV. 
While introduction of one bulk Higgs gives
rise to a more successful unification with the separation between the GUT 
scale and the compactification scale, 
KK modes of the bulk Higgs contribute flavor dependent 
power corrections to the renormalization group equations
due to the Yukawa interaction. Let us tackle the more detail of this problem 
in the section 7. 
 
In our model, we assume that
the dimension-four operators with baryon number violation on the
brane are not allowed due to $R$ parity. Moreover, when we assume that
the colored Higgsinos on the brane get masses of the GUT
scale after the GUT breaking,
there is no proton decay problem coming from the dimension-five
operators with colored Higgsino exchanges either.
Even if we had put Higgs multiplet(s) in the bulk, the effective suppression
scale for the dimension-five operators can be higher enough for the proton
longevity due to the suppression
of the wave functions of colored Higgsinos on the brane\cite{nomura}. 

On the other hand, one might worry about the potential rapid proton decay
from the exchange of the $X,Y$ gauge bosons of order $M_c\sim 10^{15}$ GeV,
which would be expected to be
the suppression scale of the dimension-six proton decay operators.
However, the coupling of the lowest KK modes of the $X,Y$ gauge
bosons to fields on the brane is suppressed
by a factor ${\rm cos}(M^A_n\pi R)\simeq \cos(\pi (n+1/2)(1-\zeta/N))$.
Therefore, the suppression factor of
the dimension-six operators becomes
\begin{eqnarray}
2\sum_{n=0}^\infty \frac{\cos^2(M^A_n\pi R)}{(M^A_n)^2}
&\simeq&8\sum_{n=0}^\infty\frac{1}{M^2_V+(2n+1)^2M^2_c}
\nonumber \\
&=&\frac{2\pi\, {\rm tanh}(\pi M_V/2M_c)}{M_V M_c}
\end{eqnarray}
where we used the eigenvalue equation (\ref{gaugeevalue}).
Then, for $M_V\gg M_c$ and $N=10^2$,
the {\it effective mass} of the X, Y gauge bosons is
$\simeq g_5 V\sqrt{M_c/2\pi}=g_4 V\simeq 0.28 V$ at the GUT scale.
The value of $V$ near the 5D fundamental scale
is sufficient to avoid the rapid proton decay from the dimension-six operators.

\section{Mass correction of the GUT scalar and softness of brane SUSY breaking}

In this section, before going into the renormalization group equations
for the soft masses, let us compute the one-loop correction to the self energy
of a GUT scalar $\phi$ located at the brane. 
There are five Feynman diagrams contributing to the self energy 
of the scalar $\phi$ as shown in Fig.~3, which are composed of four 
diagrams containing one bulk gauge field and one brane field and one diagram
involving a self interaction of the scalar. 
Since the GUT symmetry is broken on the brane, 
we should take into account the contributions from the SM
gauge fields and the $X,Y$ gauge fields separately.

Let us start with the contribution from the SM gauge fields. 
As argued in the section 3, the SM gauge bosons and gauginos have the same 
gauge couplings to the brane scalar $\phi$ without gaugino mass at $y=0$. 
When we consider the zero external momentum to see the momentum-independent
mass corrections, we obtain contributions from all the five diagrams as
\begin{eqnarray}
-i\,(m^2_\phi)^i_{j,SM}=g^2_4T^aT^a\delta^i_j
\sum_{n=-\infty}^{\infty}
\int\frac{d^4 k}{(2\pi)^2}\frac{1}{k^2-M^2_n}\frac{1}{k^2-m^2_\phi} 
N(k,M_n,m_\phi) 
\end{eqnarray}
where $i,j$ are generation indices, $g_4=g_5/\sqrt{2\pi R}$, 
$a$ runs over the SM generators, 
$M_{-n}=M_n$ and 
\begin{eqnarray}
N(k,M_n,m_\phi)&=&-k^2+4(k^2-m^2_\phi)-4(k^2-m^2_\phi)
\frac{k^2-M^2_n}{k^2-(M^{SM}_n)^2}+M^2_n+(k^2-M^2_n)
\nonumber \\
&=&4(k^2-m^2_\phi)\bigg[\frac{M^2_n-(M^{SM}_n)^2}{k^2-(M^{SM}_n)^2}\bigg].
\end{eqnarray}
Therefore, we find that there is no quadratic divergence in the integrand 
of the 4D momentum integral, which was cancelled between the five diagrams. 
With the dimensional regularization ($d=4-\epsilon$) for the 4D momentum 
integral, we obtain the mass correction to the scalar $\phi$ as
\begin{eqnarray}
(m^2_\phi)^i_{j,SM}&=&\frac{4g^2_4}{16\pi^2}T^aT^a\delta^i_j
\sum_{n=-\infty}^{\infty}\bigg[((M^{SM}_n)^2-M^2_n)
(\frac{2}{\epsilon}+\ln(4\pi)-\gamma) \nonumber \\
&-&(M^{SM}_n)^2
\ln\frac{(M^{SM}_n)^2}{\mu^2}+(M_n)^2\ln\frac{M^2_n}{\mu^2}\bigg]
\end{eqnarray} 
where $\mu$ is the renormalization scale.

On the other hand, the $X,Y$ gauge bosons and gauginos have different 
gauge couplings to the brane scalar $\phi$ even with a softness condition 
for the SM gauginos, $\varepsilon_0=0$. Therefore,  
quadratic divergences of the scalar mass would not be cancelled between 
$X,Y$ bosonic and fermionic loop diagrams unlike for the SM gauge fields. 
Likewise, with the zero external momentum, contributions involving $X,Y$ gauge
fields become 
\begin{eqnarray}
-i\, (m^2_\phi)^i_{j,X}=g^2_4 T^{\hat a}T^{\hat a}\delta^i_j
\sum_{n=-\infty}^{\infty}
\int\frac{d^4 k}{(2\pi)^2}\frac{1}{k^2-(M^A_n)^2}\frac{1}{k^2-m^2_\phi}
N(k,M^A_n,M^X_n,m_\phi)
\end{eqnarray} 
where $\hat a$ runs over broken generators, $M^A_{-n}=M^A_{n-1}$ for 
$n\geq 1$ and 
\begin{eqnarray}
N(k,M^A_n,M^X_n,m_\phi)&=&(N^2_n\cos^2\theta^A_n)[-k^2
+4(k^2-m^2_\phi)+(M^A_n)^2+(k^2-(M^A_n)^2)] \nonumber \\
&-&4(K^2_n\cos^2\theta^X_n)(k^2-m^2_\phi)
\frac{k^2-(M^A_n)^2}{k^2-(M^X_n)^2}\nonumber \\
&=&4(k^2-m^2_\phi)\bigg[N^2_n\cos^2\theta^A_n
-(K^2_n\cos^2\theta^X_n)\frac{k^2-(M^A_n)^2}{k^2-(M^X_n)^2}\bigg].
\end{eqnarray}
Consequently, using the cutoff regularization for the loop integral 
to see the divergence structure explicitly, we find that the
quadratic divergent mass correction to the scalar $\phi$ is non-vanishing 
\begin{eqnarray}
(m^2_\phi)^i_{j,X}&=&\frac{4g^2_4}{16\pi^2}T^{\hat a}T^{\hat a}
\delta^i_j \sum_{n=-\infty}^{\infty}\bigg[N^2_n\cos^2\theta^A_n 
\bigg(\Lambda^2-(M^A_n)^2\ln\frac{\Lambda^2+(M^A_n)^2}{(M^A_n)^2}\bigg) 
\nonumber \\
&-&K^2_n\cos^2\theta^X_n  
\bigg(\Lambda^2-(M^X_n)^2\ln\frac{\Lambda^2+(M^X_n)^2}{(M^X_n)^2}\bigg)\bigg].
\end{eqnarray} 
Using the mass spectrum of $X,Y$ gauge fields and considering the KK
modes below the cutoff scale, we obtain the dominant piece
of quadratically divergent mass correction as
\begin{eqnarray}
(m^2_\phi)_{X}\simeq \frac{\alpha N}{\pi^3}\bigg(\frac{M_c}{M_V}\bigg)^2
\varepsilon_\pi^2\Lambda^2\simeq (139{\rm GeV})^2
\end{eqnarray}
where we chose the cutoff scale to be of order the GUT scale 
and we used $\alpha N\simeq 0.6$ at the unification scale, 
the number of KK modes below the cutoff scale as $N=100$ and $M_c/M_V=10^{-2}$. 
Therefore, we find that due to the partial cancellation between fermionic and
bosonic contributions and the suppression of wave functions of $X,Y$ gauge 
fields at the brane, 
the resulting quadratic divergence is sufficiently softened than
the usual case without supersymmetry.
Since there is no quadratic divergence in the loop correction 
for the gaugino mass, the gaugino mass of order the weak scale is radiatively
stable, which guarantees the UV insensitivity of the one-loop scalar mass. 

Then, without worrying about the $UV$ sensitivity of the scalar mass, 
we can also obtain the one-loop correction from $X,Y$ gauge fields in the
dimensional regularization as 
\begin{eqnarray}
(m^2_\phi)^i_{j,X}&=&\frac{4g^2_4}{16\pi^2}T^{\hat a}T^{\hat a}
\delta^i_j\sum_{n=-\infty}^{\infty}\bigg[\bigg(K^2_n\cos^2\theta^X_n (M^X_n)^2
-N^2_n\cos^2\theta^A_n (M^A_n)^2\bigg)(\frac{2}{\epsilon}+\ln(4\pi)-\gamma)
\nonumber\\
&-&K^2_n\cos^2\theta^X_n (M^X_n)^2\ln\frac{(M^X_n)^2}{\mu^2}
+N^2_n\cos^2\theta^A_n (M^A_n)^2\ln\frac{(M^A_n)^2}{\mu^2}\bigg].
\end{eqnarray}

Consequently, adding all the gauge field contributions to the scalar mass,
we get the renormalization group equation for the soft mass of the scalar 
$\phi$ as 
\begin{eqnarray}
\Lambda \frac{d(m^2_\phi)^i_j}{d\Lambda}&=&-\frac{8g^2_4}{16\pi^2}\delta^i_j
\bigg[T^aT^a\sum_{|M_n|<\Lambda}((M^{SM}_n)^2-M^2_n) 
\nonumber\\
&+&T^{\hat a}T^{\hat a}\sum_{|M_n|<\Lambda}
(K^2_n\cos^2\theta^X_n (M^X_n)^2-N^2_n\cos^2\theta^A_n (M^A_n)^2)\bigg] 
\nonumber \\
&\simeq&-\frac{8g^2_4}{16\pi^2}(T^aT^a+f(\Lambda)T^{\hat a}T^{\hat a})
\delta^i_j \bigg(\frac{2\Lambda}{M_c}\bigg)|M|^2 
\end{eqnarray}
where $M$ is the lowest SM gaugino mass given by
\begin{eqnarray}
M=\frac{1}{\pi R}{\rm arctan}(\frac{\varepsilon_\pi}{2})\simeq
\frac{\varepsilon_\pi M_c}{2\pi},
\end{eqnarray}
and  
\begin{eqnarray}
f(\Lambda)=\sum_{|M_n|<\Lambda}\cos^2\theta^{A,X}_n\simeq 
1-\frac{M_V}{2\Lambda}{\rm arctan}\bigg(\frac{2\Lambda}{M_V}\bigg).
\end{eqnarray}
Here we find that the contributions coming from the KK modes 
of gauge fields add up to the {\it power running} of the scalar mass.
Since $0<f(\Lambda)<0.45$ for $M_c<\Lambda<M_V$, contributions coming 
from the $X,Y$ gauge fields are small compared to those coming from the
SM gauge fields due to the suppression of their wave functions 
at the brane.

\section{Renormalization group equations for soft masses 
and SUSY flavor problem}

The KK modes of bulk gauge fields can also give the power running 
of the other soft mass parameters which are located at the brane. 
For convenience, we assume the
boundary superpotential and the corresponding soft SUSY breaking(SSB)
Lagrangian at the $y=0$ brane in the following way, 
\begin{eqnarray}
W(\Phi)&=&\frac{1}{6}Y^{ijk}\Phi_i\Phi_j\Phi_k+\frac{1}{2}\mu_{ij}\Phi_i\Phi_j,
\\
-{\cal L}_{SSB}&=&\bigg(\frac{1}{6}h^{ijk}\phi_i\phi_j\phi_k
+\frac{1}{2}B^{ij}\phi_i\phi_j+h.c.\bigg)+\phi^{*j}(m^2)^i_j\phi_j.
\end{eqnarray} 
Here let us remind that there exists a gaugino mass term only 
at the $y=\pi R$ brane. 

Upon the orbifold compactification on $S^1/Z_2$, 
only the even modes of bulk fields are coupled 
to the brane\footnote{The odd modes can also couple to the brane 
only with their derivatives with respect to the extra dimension.}, 
so each KK mode gives a logarithmic contribution to two and three point
functions for brane fields as in 4D\cite{DDG,kubo1}. 
Therefore, logarithmic contributions 
to the anomalous dimension of a brane field from the KK modes below the cutoff 
scale sum up to the power running. 
For a Higgs field in the bulk,  
the anomalous dimension of its zero mode gets a 4D logarithmic contribution
from brane fields  
due to Yukawa interactions at the brane while it does not have a loop 
contribution from the bulk gauge fields 
thanks to the bulk ${\cal N}=2$ supersymmetry\cite{DDG,kubo1}.
For applicability to the case with extra dimension, let us recall
conventional formulae for one-loop beta functions of couplings
in 4D\cite{jones} as follows
\begin{eqnarray} 
\beta_g&=&\frac{g^3}{16\pi^2}[\sum_i l(R_i)-3C_2(G)], \\
\beta_M&=&\frac{2g^2}{16\pi^2}[\sum_i l(R_i)-3C_2(G)] M, \\
\beta^{ijk}_Y &=&\sum_l Y^{ijl}\gamma^k_l+(k\leftrightarrow i)
+(k\leftrightarrow j), \\
\beta^{ij}_\mu&=&\gamma^i_l\mu^{lj}+\gamma^j_l\mu^{il}, \\
\beta^{ij}_B&=&\gamma^i_l B^{lj}+\gamma^j_l B^{il}-2\gamma^i_{1l}\mu^{lj}
-2\gamma^j_{1l}\mu^{il}, \\
\beta^{ijk}_h&=&\gamma^i_l h^{ljk}+\gamma^j_l h^{ilk}+\gamma^k_l h^{ijl}
-2\gamma^i_{1l}Y^{ljk}-2\gamma^j_{1l}Y^{ilk}-2\gamma^k_{1l}Y^{ijl}, \\
(\beta_{m^2})^i_j&=&[2{\cal O}{\cal O}^*+2|M|^2g^2\frac{\partial}{\partial g^2}
+{\tilde Y}_{lmn}\frac{\partial}{\partial Y_{lmn}}
+{\tilde Y}^{lmn}\frac{\partial}{\partial Y^{lmn}}]\gamma^i_j, 
\end{eqnarray}
where $\beta_g=\Lambda \frac{d g}{d\Lambda}$ and etc., 
$\gamma^i_j$ is the anomalous dimension given by
\begin{eqnarray}
\gamma^i_j&=&\frac{1}{16\pi^2}\bigg[\frac{1}{2}\sum_{m,n}Y^{imn}Y_{jmn}
-2\delta^i_j g^2 C_2(R_i)\bigg], \\
{\cal O}&=&Mg^2\frac{\partial}{\partial g^2}
-h^{lmn}\frac{\partial}{\partial Y^{lmn}}, \\
{\tilde Y}^{lmn}&=&(m^2)^l_k Y^{kmn}+(m^2)^m_k Y^{lkn}+(m^2)^n_k Y^{lmk},
\end{eqnarray}
$\gamma^i_{1j}={\cal O}\gamma^i_j$ and $Y_{lmn}=(Y^{lmn})^*$. 
Here $l(R_i)$ denotes the index of the representation $R_i$, and 
$C_2(G)$ and $C_2(R_i)$ are the quadratic Casimirs of the adjoint
representation and the representation $R_i$ of the gauge group $G$,
respectively. 

Now let us take the case with only gauge fields in the bulk for the running
of couplings.
Then, in the flavor bases where gaugino couplings are diagonal, bulk gauge 
corrections give rise to only the diagonal 
elements of anomalous dimensions. 
Above the GUT scale, we should take into account the KK modes of bulk gauge
fields for the case without the GUT breaking. 
On the other hand, below the GUT scale
down to the compactification scale,
we have to include the GUT breaking effect to the KK modes of bulk gauge 
fields. 
Taking into account additional KK contributions to the 4D beta functions 
for soft parameters at each KK threshold, 
we find the following approximate one-loop renormalization group 
equations(RGEs)\footnote{Note that the beta functions for the
gauge coupling
and the gaugino mass are the half those in Ref.~\cite{kubo}.
They disregarded the reduction of the number of KK modes on $S^1/Z_2$.} 
for $\Lambda>M_{GUT}$ and $M_c<\Lambda<M_{GUT}$:
\begin{eqnarray}
16\pi^2\beta_g
=-2C_2(G)\bigg(\frac{\Lambda}{M_c}\bigg)g^3,\qquad\qquad \qquad \qquad 
\end{eqnarray}
\begin{eqnarray}
16\pi^2\beta_M
=-4C_2(G)\bigg(\frac{\Lambda}{M_c}\bigg)g^2 M,\qquad \quad\qquad \qquad 
\end{eqnarray}
\begin{eqnarray}
16\pi^2\beta^{ijk}_Y=\left\{
\begin{array}{l} -2(C_2(R_i)+C_2(R_j)+C_2(R_k))\left(\frac{2\Lambda}{M_c}\right)g^2 Y^{ijk},\\
-2\Big[C_2(R_i)+C_2(R_j)+C_2(R_k) \\ 
+(C^X_2(R_i)+C^X_2(R_j)+C^X_2(R_k))(f(\Lambda)-1)\Big]\left(\frac{2\Lambda}{M_c}\right)g^2 Y^{ijk},
\end{array} \right. \qquad
\end{eqnarray}
\begin{eqnarray}
16\pi^2\beta^{ij}_\mu=\left\{
\begin{array}{l} 
-2(C_2(R_i)+C_2(R_j))\left(\frac{2\Lambda}{M_c}\right)g^2\mu^{ij},\\
-2\Big[C_2(R_i)+C_2(R_j) \\ 
+(C^X_2(R_i)+C^X_2(R_j))(f(\Lambda)-1)\Big]\left(\frac{2\Lambda}{M_c}\right)g^2\mu^{ij},
\end{array} \right.\qquad\qquad\qquad \quad
\end{eqnarray}
\begin{eqnarray}
16\pi^2\beta^{ij}_B=\left\{
\begin{array}{l}
 2(C_2(R_i)+C_2(R_j))\left(\frac{2\Lambda}{M_c}\right)g^2(2M\mu^{ij}-B^{ij}), \\
2\Big[C_2(R_i)+C_2(R_j) \\
+(C^X_2(R_i)+C^X_2(R_j))(f(\Lambda)-1)\Big]\left(\frac{2\Lambda}{M_c}\right)g^2(2M\mu^{ij}-B^{ij}),
\end{array} \right. \qquad 
\end{eqnarray} 
\begin{eqnarray}
16\pi^2\beta^{ijk}_h=\left\{
\begin{array}{l}
 2(C_2(R_i)+C_2(R_j)+C_2(R_k))\left(\frac{2\Lambda}{M_c}\right)g^2 
(2MY^{ijk}- h^{ijk}),\\
2\Big[C_2(R_i)+C_2(R_j)+C_2(R_k) \\ 
+(C^X_2(R_i)+C^X_2(R_j)+C^X_2(R_k))(f(\Lambda)-1)\Big]
\left(\frac{2\Lambda}{M_c}\right)g^2 (2MY^{ijk}-h^{ijk}),
\end{array} \right. 
\end{eqnarray}
\begin{eqnarray} 
16\pi^2(\beta_{m^2})^i_j=\left\{
\begin{array}{l}
  -8\delta^i_j C_2(R_i)\left(\frac{2\Lambda}{M_c}\right)g^2|M|^2,\\
-8\delta^i_j \Big[C_2(R_i)+C^X_2(R_i)(f(\Lambda)-1)\Big] 
\left(\frac{2\Lambda}{M_c}\right)g^2|M|^2
\end{array} \right.\qquad \quad \qquad \qquad
\end{eqnarray}
where 
$C^X_2(R_i)$ is the broken part of the quadratic Casimir of the 
representation $R_i$. Note that $C^X_2(R_i)$ is different for the different 
representations of the unbroken gauge group in the same GUT multiplet. 
Therefore, for the nonzero 
factor $(f(\Lambda)-1)$, 
the runnings of soft mass parameters for the same GUT multiplet become
non-universal between the GUT scale and the compactification scale. 

Then, we find that the RGEs for $M$, $Y^{ijk}$ and $\mu^{ij}$ 
for $M_c<\Lambda<M_{GUT}$ are solved by
\begin{eqnarray}
M(M_c)&=&\bigg(\frac{g(M_c)}{g(M_{GUT})}\bigg)^2 M(M_{GUT}), \label{relg1}\\
Y^{ijk}(M_c)&=&\bigg(\frac{g(M_c)}{g(M_{GUT})}\bigg)^{2\eta^{ijk}}
\bigg(\frac{g(M_{GUT})}{g(M_c)}\bigg)^{2t_1\eta^{ijk}_X} Y^{ijk}(M_{GUT}), 
\label{relg2}\\
\mu^{ij}(M_c)&=&\bigg(\frac{g(M_c)}{g(M_{GUT})}\bigg)^{2\eta^{ij}}
\bigg(\frac{g(M_{GUT})}{g(M_c)}\bigg)^{2t_1\eta^{ij}_X} \mu^{ij}(M_{GUT}),
\label{relg3}
\end{eqnarray} 
where 
\begin{eqnarray}
\eta^{ijk}&=&\frac{C_2(R_i)+C_2(R_j)+C_2(R_k)}{C_2(G)}, \\
\eta^{ij}&=&\frac{C_2(R_i)+C_2(R_j)}{C_2(G)}, \\
t_1&=&\bigg[\ln\frac{g(M_c)}{g(M_{GUT})}\bigg]^{-1}\int^{g(M_{GUT})}_{g(M_c)}
(f(\Lambda)-1)\frac{dg}{g},
\end{eqnarray}
and $\eta^{ijk}_X$, $\eta^{ij}_X$ are the broken part of $\eta^{ijk}$ 
and $\eta^{ij}$, respectively. For instance, in the $SU(5)$ case, 
$\eta^{ijk}=48/25(42/25)$ for the up(down) type Yukawa coupling and 
$\eta^{ij}=24/25$. Note also that
$\eta^{ijk}_X=1(4/5),6/5$ for the up(down) type quark Yukawa coupling and 
the lepton Yukawa coupling in order while $\eta^{ij}_X=3/5(2/5)$ for the
Higgs doublet(triplet). So, using the values, $\alpha(M_c)\simeq\frac{1}{23}$
for $M_c\simeq 10^{15}$ GeV, 
$g(M_c)/g(M_{GUT})\simeq 2.8$ and $t_1=0.554$, 
the up type quark Yukawa coupling 
at $M_c$ is lowered by 0.692 than the case without GUT breaking
and the down type quark and lepton Yukawa couplings at $M_c$ are different 
with $Y_d:Y_l=0.404 : 0.257$. 
The difference between $\mu$ parameters for the
Higgs doublet and triplet at $M_c$ comes with  
$\mu_{H_u}(=\mu_{H_d}):\mu_{H_C}(=\mu_{{\bar H}_C})=0.507 : 0.692$.  
Therefore, the GUT breaking effect
only gives rise to the $O(1)$ difference
between Yukawa couplings (or $\mu$ terms) in the same GUT multiplet at
the compactification scale. 
However, compared to the case that the GUT scale is the same 
as the compactification scale\cite{kubo}, we find that
the overall magnitude of Yukawa coupling and the $\mu$ term becomes 
even larger at $M_c$
due to the difference between the GUT scale and the compactification scale.

On the other hand, the ratios for the SSB parameters $B^{ij}$, $h^{ijk}$ 
and $(m^2)^i_j$ to the gaugino mass 
have the infrared fixed points shifted due to the GUT breaking effect compared
to the case without the GUT breaking\cite{kubo}. 
We also obtain the explicit solutions for those SSB parameters as
\begin{eqnarray}
\frac{B^{ij}}{M\mu^{ij}}(M_c)&=&-2(\eta^{ij}-t_2\eta^{ij}_X)
+\bigg(\frac{g(M_{GUT})}{g(M_c)}\bigg)^2
\bigg(\frac{B^{ij}}{M\mu^{ij}}(M_{GUT})+2\eta^{ij}\bigg), \label{irf1}\\
\frac{h^{ijk}}{MY^{ijk}}(M_c)&=&-2(\eta^{ijk}-t_2\eta^{ijk}_X)
+\bigg(\frac{g(M_{GUT})}{g(M_c)}\bigg)^2
\bigg(\frac{h^{ijk}}{MY^{ijk}}(M_{GUT})+2\eta^{ijk}\bigg), \label{irf2}\\
\frac{(m^2)^i_j}{|M|^2}(M_c)&=&\frac{2(C_2(R_i)-t_3 C^X_2(R_i))}{C_2(G)}
\delta^i_j \nonumber \\
&+&\bigg(\frac{g(M_{GUT})}{g(M_c)}\bigg)^4
\bigg(\frac{(m^2)^i_j}{|M|^2}(M_{GUT})-\frac{2C_2(R_i)}{C_2(G)}\delta^i_j\bigg)
\label{irf3}
\end{eqnarray} 
where
\begin{eqnarray}
t_2&=&\frac{2}{g^2(M_c)}\int^{g(M_{GUT})}_{g(M_c)}g^2
(f(\Lambda)-1)\frac{dg}{g}, \\
t_3&=&\frac{4}{g^4(M_c)}\int^{g(M_{GUT})}_{g(M_c)}g^4
(f(\Lambda)-1)\frac{dg}{g}.
\end{eqnarray}
Therefore, the diagonal elements of the soft scalar masses become dominant and 
degenerate.
Moreover, inserting the following approximate value for the ratio of the gauge
couplings into eq.~(\ref{irf3}) for $G=SU(5)$ 
\begin{eqnarray}
\frac{g(M_c)}{g(M_{GUT})}\simeq \bigg[\frac{C_2(G)\alpha(M_c)}{\pi}\bigg]^{1/2}
\bigg(\frac{M_{GUT}}{M_c}\bigg)^{1/2}\simeq 2.63,
\end{eqnarray}
where we used $\alpha(M_c)\simeq \frac{1}{23}$ for $M_c\simeq 10^{15}$ GeV 
and $M_{GUT}/M_c=10^2$,
the $O(1)$ off-diagonal components of $(m^2)^i_j/|M|^2$ initially given 
at $M_{GUT}$ can be suppressed to be $O(10^{-2})$ as shown in the second term
of eq.~(\ref{irf3}). So, for the contribution of the soft masses to 
$(\delta_{ii})_{LL,RR}$, we can satisfy the stringent bounds from  
the $K_L-K_S$ mass difference $\Delta m_K$ and the decay
$\mu\rightarrow e\gamma$\cite{masiero}.
Likewise, from eq.~(\ref{irf2}), when the size of $h^{ijk}$ is assumed 
to be of order $MY^{ijk}$ at $M_{GUT}$, we also obtain the
non-aligned components of the trilinear couplings as follows
\begin{eqnarray}
\bigg|\frac{h^{ijk}}{M}(M_c)+2(\eta^{ijk}
-t_2\eta^{ijk}_X)Y^{ijk}(M_c)\bigg|\simeq
\bigg(\frac{g(M_{GUT})}{g(M_c)}\bigg)^2 O(Y^{ijk}(M_c))
\end{eqnarray}
Thus, in the $SU(5)$ case,
we find that the non-aligned components of the trilinear couplings contribute
to $(\delta_{ij})_{LR}$:
$(\delta^{l(d)}_{ij})_{LR}\simeq \xi \langle H_d \rangle 
MY^{l(d)}_{ij}/m^2_{{\tilde l}({\tilde q})}$
for the sleptons(down squarks) and
$(\delta^u_{ij})_{LR}\simeq \xi\langle H_u \rangle 
MY^{u}_{ij}/m^2_{\tilde q}$ 
for the up squarks where $\xi\equiv (g(M_{GUT})/g(M_c))^2$. 
Therefore, in the basis that the up quark mass matrix
is diagonal\cite{masiero2}, we obtain the LR mass terms contributing to
$\epsilon'/\epsilon$, $b \rightarrow s \gamma$  
and $\mu\rightarrow e \gamma$ decay processes in order as
${\rm Im}(\delta^{d}_{12})_{LR}\simeq \xi m_sV_{us}M/m^2_{\tilde q}\sim 2.9\times 10^{-6} $, 
$(\delta^{d}_{23})_{LR}\simeq \xi m_b V_{cb}M/m^2_{\tilde q}\sim 2.9\times 10^{-6}$ 
and $(\delta^l_{12})_{LR}\simeq \xi m_\mu V_{\nu_e\mu}M/m^2_{\tilde l}\sim 3.0\times 10^{-5} V_{\nu_e\mu}$ 
where $V_{us}$, $V_{cb}$ and $V_{\nu_e\mu}$ 
denote quark and lepton mixings. Henceforth 
we take $m_{\tilde q}\sim m_{\tilde l}\sim 500{\rm GeV}$
and assume that the photino(gluino)
mass is the same as the slepton(squark) mass. 
Therefore, we can satisfy or saturate 
the experimental limits on the LR mass terms such as
${\rm Im}(\delta^{d}_{12})_{LR}<2.1\times 10^{-5}$,
$(\delta^{d}_{23})_{LR}<1.6\times 10^{-2}$,
and $(\delta^l_{12})_{LR}<4.3\times 10^{-6}$.

The diagonal terms of trilinear couplings with the nonzero phases
also contribute to
${\rm Im}(\delta_{ii})_{LR}$\cite{masiero}. 
The most stringent contraints on those come from the electric dipole moments
of the neutron and the electron\cite{masiero}:
${\rm Im}(\delta^l_{11})_{LR}<9.3\times 10^{-6}$ and
${\rm Im}(\delta^d_{11})_{LR}<3.0\times 10^{-6}$.
Since the phases of $h^{ijk}/MY^{ijk}$
given at $M_{GUT}$ are suppressed to be of order $10^{-2}$ 
due to the power running from eq.~(\ref{irf2}), we obtain the sufficient 
suppression of LR mass terms to satisfy the EDM constraints 
as ${\rm Im}(\delta^l_{11})_{LR}\sim \xi M m_e/m^2_{\tilde l}\sim 1.5\times 10^{-7}$ and
${\rm Im}(\delta^d_{11})_{LR}\sim \xi M m_d/m^2_{\tilde q}\sim 4.4\times 10^{-7}$.
Likewise, the phases of B terms are sufficiently suppressed at $M_c$. 
On the other hand, in the model with one bulk Higgs 
multiplet, which is needed for a successful unification of gauge couplings, 
the large difference of scales, $M_{GUT}$ and $M_c$, is compatible  
with the successful unification of gauge couplings and the proton longevity.
However, the flavor violation due to 
power-law Yukawa interaction could make the soft parameters non-universal. 
In the next section, let us deal with 
the power-law Yukawa terms giving rise to the flavor violation again. 

For $\Lambda$ close to $M_c$, the logarithmic corrections will not be 
negligible, which could destroy the universality of the SSB terms. 
However, in the concrete example of the SU(5) GUT, it has been shown that the 
logarithmic corrections coming from the Yukawa interactions are small enough 
to avoid the harmful flavor or $CP$ violating processes\cite{kubo}. 

For concreteness, let us consider the infrared fixed points of the soft mass
parameters in the $SU(5)$ model. The SSB Lagrangian except for the gauginos 
is given by
\begin{eqnarray}
-{\cal L}_{SSB}&=&m^2_{H}H^\dagger H+m^2_{\bar H}{\bar H}^\dag {\bar H}
+m^2_\Sigma{\rm tr}(\Sigma^\dagger \Sigma) \nonumber \\
&+&\sum_{i,j}^3[(m^2_{\Psi({\bar 5})})^{ij}\Psi_i({\bar 5})^\dag
\Psi_j({\bar 5})+(m^2_{\Psi(10)})^{ij}{\rm tr}
(\Psi_i(10)^\dag\Psi_j(10))] \nonumber \\
&+&\bigg\{B_\Sigma {\rm tr} \Sigma^2+\frac{h_\lambda}{3}{\rm tr} \Sigma^3
+h_f {\bar H}\Sigma H +B_H {\bar H}H \nonumber \\
&+&\frac{h^{ij}_U}{4}\Psi_i(10)\Psi_j(10)H 
+\sqrt{2}h^{ij}_D\Psi_i({\bar 5})\Psi_j(10){\bar H}+h.c.\bigg\}
\end{eqnarray} 
where it is understood that all fields are scalar components 
of the corresponding superfields. Then, due to the GUT breaking effect, 
the soft mass parameters for the same GUT multiplet have different infrared 
fixed points. Using the general formulae, eqs.~(\ref{irf1})-(\ref{irf3}), 
and the numerical values $t_2=0.479$ and $t_3=0.544$, 
the SSB parameters approach the following values 
in the limit of
the energy scale going down to the compactification scale:
\begin{eqnarray}
B_{H_u},B_{H_d}&\rightarrow& -\frac{48}{25}M\mu_{H_u}(1-\frac{5}{8}t_2)
=-1.35M\mu_{H_u}, \\
B_{H_C},B_{H_{\bar C}}&\rightarrow& -\frac{48}{25}M\mu_{H_C}(1-\frac{5}{12}t_2)
=-1.54M\mu_{H_C}, \\
h_U&\rightarrow&-\frac{96}{25}MY_U(1-\frac{25}{48}t_2)
=-2.88MY_U, \\
h_d&\rightarrow&-\frac{84}{25}MY_d(1-\frac{10}{21}t_2)
=-2.59MY_d, \\
h_l&\rightarrow&-\frac{84}{25}MY_l(1-\frac{5}{7}t_2)
=-2.21MY_l, \\
m^2_{\tilde q}&\rightarrow& \frac{36}{25}|M|^2(1-\frac{5}{12}t_3)
=1.11|M|^2, \\
m^2_{{\tilde u}^c}&\rightarrow& \frac{36}{25}|M|^2(1-\frac{5}{9}t_3)
=1.01|M|^2, \\
m^2_{{\tilde e}^c}&\rightarrow& \frac{36}{25}|M|^2(1-\frac{5}{6}t_3)
=0.788|M|^2, \\
m^2_{{\tilde d}^c},m^2_{H_C},m^2_{H_{\bar C}}&\rightarrow& 
\frac{24}{25}|M|^2 (1-\frac{5}{12}t_3)=0.742|M|^2,\\
m^2_{\tilde l},m^2_{H_u},m^2_{H_d}&\rightarrow&
\frac{24}{25}|M|^2(1-\frac{5}{8}t_3)=0.634|M|^2, 
\end{eqnarray}
where we considered soft parameters in terms of components of the GUT multiplet 
under the SM group representations. Note that at the GUT scale, 
$\mu_{H_u}=\mu_{H_C}=\mu_H$,
$B_{H_u}=B_{H_C}=B_{H}$, $B_{H_d}=B_{H_{\bar C}}=B_{\bar H}$,
$h_d=h_l=h_D$, $Y_d=Y_l=Y_D$,
$m^2_{\tilde q}=m^2_{{\tilde u}^c}=m^2_{{\tilde e}^c}=m^2_{\Psi(10)}$,
$m^2_{{\tilde d}^c}=m^2_{\tilde l}=m^2_{\Psi({\bar 5})}$, $m^2_{H_u}=m^2_{H_C}=m^2_{H}$, 
and $m^2_{H_d}=m^2_{H_{\bar C}}=m^2_{\bar H}$.

\section{Flavor universality with bulk Higgs multiplet(s)}

In the section 4, we have shown that letting Higgs fields propagate in the bulk 
is needed to get a more successful unification of gauge couplings 
and it is also possible to have
a large separation between the GUT scale and the compactification scale. 
However, introduction of bulk Higgs fields would lead to a power-law
contribution of additional KK modes to the RGEs, which does not respect the 
flavor universality. Moreover, as mentioned in the previous section, 
there is no gauge correction to the anomalous dimension of the zero mode 
of a bulk Higgs field 
due to the ${\cal N}=2$ non-renormalization theorem. Therefore, in this section,
we will present the RGEs for soft parameters in the case with bulk
Higgs field(s) and show how infrared fixed points of soft parameters 
can be maintained.  

In the case with bulk Higgs field(s)\footnote{Note that we have to put a SUSY 
breaking mass for bulk Higgs scalar(s) at $y=\pi R$ for softness as for bulk gauginos. Then, the brane 
SUSY breaking parameter for bulk Higgs scalar(s) has a mass dimension of one. 
Of course, a brane Higgs has a soft mass at $y=0$.}, 
we use the 4D results to obtain the 
anomalous dimensions for the brane matters $\bf 10$, $\bf {\bar 5}$ and bulk
Higgs field(s)\cite{kubo1}. 
Here we consider the case (1) with the down type Higgs in the bulk 
and the up type Higgs 
on the brane and the case (2) with both two Higgs fields in the bulk.
Then, when we neglect the $O(1)$ GUT breaking effect for the time being, 
we can get one-loop anomalous dimensions above the compactification scale 
with the upper and lower ones in the braces for the cases (1) and (2), 
respectively, 
\begin{eqnarray} 
16\pi^2 \gamma_{10}&=&\left\{
\begin{array}{l}
(-\frac{36}{5}g^2+2Y_D Y_D^\dag )\bigg(\frac{2\Lambda}{M_c}\bigg)+3Y_U Y_U^\dag
\\
(-\frac{36}{5}g^2
+3Y_U Y_U^\dag+2Y_D Y_D^\dag )\bigg(\frac{2\Lambda}{M_c}\bigg), 
\end{array} \right. \label{ad1}\\
16\pi^2 \gamma_{\bar 5}&=&(-\frac{24}{5}g^2+4Y_D Y_D^\dag )
\bigg(\frac{2\Lambda}{M_c}\bigg), \label{ad2}\\ 
16\pi^2 \gamma_H&=&\left\{
\begin{array}{l}
-\frac{24}{5}g^2\bigg(\frac{2\Lambda}{M_c}\bigg)+3{\rm Tr}Y_U Y_U^\dag \\
3 {\rm Tr}Y_U Y_U^\dag, 
\end{array} \right.\label{ad3} \\
16\pi^2 \gamma_{\bar H}&=& 4{\rm Tr}Y_D Y_D^\dag.\label{ad4}
\end{eqnarray}
Note that there are additional power-law corrections 
to the anomalous dimensions of
brane matters $\bf 10$, $\bf {\bar 5}$ due to the KK modes 
of bulk Higgs multiplet(s). On the other hand,  
the zero mode of a bulk Higgs field 
only obtains a logarithmic anomalous dimension from the brane 
matters via the Yukawa couplings 
as shown in the lower one of eq.~(\ref{ad3}) and eq.~(\ref{ad4}). 

Consequently, using the above anomalous dimensions and the 4D beta functions
and keeping only the dominant part of one-loop beta functions, we obtain those
for the rigid supersymmetry parameters as 
\begin{eqnarray}
16\pi^2\beta_g&=&\left\{
\begin{array}{l}
-9\bigg(\frac{\Lambda}{M_c}\bigg)g^3, \\
-8\bigg(\frac{\Lambda}{M_c}\bigg)g^3, 
\end{array} \right. \label{rg1}\\
16\pi^2\beta_M&=&\left\{
\begin{array}{l}
-18\bigg(\frac{\Lambda}{M_c}\bigg)g^2 M, \\
-16\bigg(\frac{\Lambda}{M_c}\bigg)g^2 M, 
\end{array} \right. \label{rg2}\\
16\pi^2\beta_{Y_U}&=&\left\{
\begin{array}{l}
(-\frac{96}{5}g^2+4Y_D Y_D^\dag )Y_U\bigg(\frac{2\Lambda}{M_c}\bigg), \\
(-\frac{72}{5}g^2+6Y_U Y_U^\dag+4Y_D Y_D^\dag )Y_U
\bigg(\frac{2\Lambda}{M_c}\bigg), 
\end{array} \right. \label{rg3}\\
16\pi^2\beta_{Y_D}&=&\left\{
\begin{array}{l}
(-12g^2+6Y_D Y_D^\dag )Y_D
\bigg(\frac{2\Lambda}{M_c}\bigg), \\
(-12g^2+3{\rm Tr}Y_U Y_U^\dag+6Y_D Y_D^\dag )Y_D
\bigg(\frac{2\Lambda}{M_c}\bigg), 
\end{array} \right. \label{rg4}\\
16\pi^2\beta_\mu&=&\left\{
\begin{array}{l}
-\frac{24}{5}g^2\bigg(\frac{2\Lambda}{M_c}\bigg)\mu, \\
(3{\rm Tr}Y_U Y_U^\dag+4{\rm Tr}Y_D Y_D^\dag )\mu, 
\end{array} \right. \label{rg5}
\end{eqnarray}
and those for the soft supersymmetry breaking parameters as 
\begin{eqnarray}
16\pi^2\beta_B&=&\left\{
\begin{array}{l}
\frac{24}{5}g^2\bigg(\frac{2\Lambda}{M_c}\bigg)(2\mu M- B), \\
(3{\rm Tr}Y_U Y_U^\dag+4{\rm Tr}Y_D Y_D^\dag )B+ (6{\rm Tr}h_UY_U^\dag
+8{\rm Tr}h_D Y_D^\dag )\mu, 
\end{array} \right. \label{soft1}\\
16\pi^2\beta_{h_U}&=&\left\{
\begin{array}{l}
\bigg[\frac{96}{5}g^2(2MY_U-h_U)+4Y_D Y_D^\dag h_U
+8({\rm Tr}h_DY_D^\dag)Y_U\bigg]\bigg(\frac{2\Lambda}{M_c}\bigg), \\
\bigg[\frac{72}{5}g^2(2MY_U-h_U)+(6Y_U Y_U^\dag+4Y_D Y_D^\dag)h_U \\
\qquad+(12{\rm Tr}h_U Y_U^\dag+8{\rm Tr}h_DY_D^\dag)Y_U\bigg]
\bigg(\frac{2\Lambda}{M_c}\bigg)\mu, 
\end{array} \right. \label{soft2}\\
16\pi^2\beta_{h_D}&=&\left\{ 
\begin{array}{l}
\bigg[12g^2(2MY_D-h_D)+6Y_D Y_D^\dag h_D+12({\rm Tr}h_DY_D^\dag) Y_D\bigg]
\bigg(\frac{2\Lambda}{M_c}\bigg), \\
\bigg[12g^2(2MY_D-h_D)+(3Y_U Y_U^\dag+6Y_D Y_D^\dag)h_D \\
\qquad+(6{\rm Tr}h_U Y_U^\dag+12{\rm Tr}h_DY_D^\dag)Y_D\bigg]
\bigg(\frac{2\Lambda}{M_c}\bigg)\mu, 
\end{array} \right. \label{soft3} 
\end{eqnarray}
\begin{eqnarray}
16\pi^2\beta_{m^2_{10}}&=&\left\{
\begin{array}{l}
\bigg[-\frac{144}{5}g^2|M|^2+4h_D h_D^\dag+4(m^2_{10}+m^2_{\bar 5}
+m^2_{H_d})Y_D Y_D^\dag\bigg]\bigg(\frac{2\Lambda}{M_c}\bigg), \\
\bigg[-\frac{144}{5}g^2|M|^2+6h_U h_U^\dag+4h_D h_D^\dag 
+6(2m^2_{10}+m^2_{H_u})Y_U Y_U^\dag \\
\qquad+4(m^2_{10}+m^2_{\bar 5}
+m^2_{H_d})Y_D Y_D^\dag\bigg]\bigg(\frac{2\Lambda}{M_c}\bigg), 
\end{array} \right. \label{soft4}\\
16\pi^2\beta_{m^2_{\bar 5}}&=&
\bigg[-\frac{96}{5}g^2|M|^2+8h_D h_D^\dag+8(m^2_{10}+m^2_{\bar 5}
+m^2_{H_u})Y_D Y_D^\dag\bigg]\bigg(\frac{2\Lambda}{M_c}\bigg), \label{soft5} 
\end{eqnarray}
\begin{eqnarray}
16\pi^2\beta_{m^2_{H_u}}&=&\left\{
\begin{array}{l}
-\frac{96}{5}g^2|M|^2\bigg(\frac{2\Lambda}{M_c}\bigg), \\
6{\rm Tr}h_U h_U^\dag+6(2m^2_{10}+m^2_{H_u}){\rm Tr}Y_U Y_U^\dag, 
\end{array} \right. \label{soft6}\\
16\pi^2\beta_{m^2_{H_d}}&=&8{\rm Tr}h_D h_D^\dag+8(m^2_{10}+m^2_{\bar 5}
+m^2_{H_d}){\rm Tr}Y_D Y_D^\dag.\label{soft7}
\end{eqnarray}
Note that only for the case (1) with the down type Higgs field in the bulk, 
we can retain the power law contribution to the beta function 
of the $B$ term or the $\mu$ term, so that there does exist 
an IR fixed point for the $B$ term which lead to a suppressed phase of $B$. 
More importantly, for the case (1), additional power terms coming only from
the down type Yukawa coupling can be negligible for universality 
of soft masses at the compactification scale. 
On the other hand, for the case (2) with both Higgs fields in the bulk,
the power running due to the up type Yukawa coupling could be 
of the same order of magnitude as that due to the gauge interaction, 
for instance, near the compactification scale. Therefore, we cannot ignore the 
potential non-universality due to the up type Yukawa coupling in the case (2).
Therefore, henceforth let us concentrate 
on the case (1), i.e. the upper one of each pair of beta functions.

Even with one bulk Higgs, we find that the gauge coupling has the power running 
with asymptotic freedom and a standard relation 
between gauge coupling and gaugino mass holds. Moreover, since there is one
Higgs field at the brane, $\mu$, $B$ terms and the soft mass 
of the up type Higgs scalar have power contributions 
only from KK modes of gauge fields. On the other hand, 
Yukawa couplings and other soft mass
parameters have the power contributions due to the KK modes of the bulk 
Higgs field via the down type Yukawa interactions, 
which are flavor dependent. However, we can show that the power-like disorder 
arising from Yukawa couplings can be negligible due to the smallness of the
down type Yukawa couplings. 

To begin with, let us consider the disorder in the Yukawa couplings due to the
KK modes of the down type bulk Higgs. For simplicity, let us neglect the 
$O(1)$ GUT breaking effect. By using eq.~(\ref{rg1}), 
we can rewrite eqs.~(\ref{rg3}) and (\ref{rg4}) as
\begin{eqnarray}
16\pi^2\Lambda \frac{d}{d\Lambda}{\tilde Y}_U
&=&4Y_D Y_D^\dag {\tilde Y}_U \bigg(\frac{2\Lambda}{M_c}\bigg), \label{yukau}\\
16\pi^2\Lambda \frac{d}{d\Lambda}{\tilde Y}_D
&=&6Y_D Y_D^\dag {\tilde Y}_D \bigg(\frac{2\Lambda}{M_c}\bigg), \label{yukad} 
\end{eqnarray}   
where 
\begin{eqnarray}
{\tilde Y}_U\equiv \frac{Y_U}{g^{2\eta_U}}, \ \ \ 
{\tilde Y}_D\equiv \frac{Y_D}{g^{2\eta_D}}
\end{eqnarray} 
with $\eta_U=32/15$ and $\eta_D=4/3$. Then, inserting the solutions 
for Yukawa couplings without Yukawa power corrections such as 
\begin{eqnarray}
{\tilde Y}_U(\Lambda)={\tilde Y}_U(M_c), \ \ \ 
{\tilde Y}_D(\Lambda)={\tilde Y}_D(M_c), \label{yukabulk}
\end{eqnarray}
into the right-hand sides of eqs.~(\ref{yukau}) 
and (\ref{yukad}), and using $Y_D/Y_U\simeq 1/60$ with $Y_U\simeq 1$ at $M_c$ 
and $\alpha(M_c)\simeq \frac{1}{23}$ for $M_c\simeq 10^{15}$ GeV, 
we can estimate the power-like disorder in the Yukawa couplings as
\begin{eqnarray}
{\tilde Y}^{-1}_U(M_c){\tilde Y}_U(\Lambda)-1
&\lesssim& \frac{4g^{-2}(M_c)}{9(2\eta_D-1)}(Y^{-1}_U Y_D Y_D^\dag Y_U)(M_c)
\sim 10^{-4}, \\
{\tilde Y}^{-1}_D(M_c){\tilde Y}_D(\Lambda)-1
&\lesssim&\frac{2g^{-2}(M_c)}{3(2\eta_D-1)}(Y_D^\dag Y_D)(M_c)
\sim 10^{-4}.
\end{eqnarray}
On the other hand, the power contribution to the $\mu$ term comes only from
the gauge interaction as can be seen in the upper one of eq.~(\ref{rg5}). 
Thus, we get the similar running of the $\mu$ term with
a different exponent of $g$ compared to the case without bulk Higgs: 
$\eta^{ij}=24/25\rightarrow \eta_\mu=2C_2(5)/9=8/15$ 
in eq.~(\ref{relg3}). 

Before going into the evolution of the soft mass parameters of brane matters,
let us consider the RG flow of the down type Higgs mass. 
The zero mode of the down type bulk Higgs only has the nonzero anomalous 
dimension via the Yukawa coupling with the brane matters 
as shown in eq.~(\ref{soft7}).
Now we show how the running of the down type Higgs 
mass can be neglected above the compactification scale 
by taking its ratio to the Yukawa couplings. 
Here we do not include the GUT
breaking effect to show the gross behavior.
Using eq.~(\ref{soft7}) and the upper one of eq.~(\ref{rg4}) and recalling  
that the dominant running masses of the brane scalars
are of order the gaugino mass,
we obtain the following approximate running equation
for ${\tilde m}^2_{H_d}=m^2_{H_d}/({\rm Tr}Y_DY_D^\dag)$
\begin{eqnarray}
16\pi^2\Lambda\frac{d}{d\Lambda}{\tilde m}^2_{H_d}
&\simeq& 8(1+a+b)|M|^2+8m^2_{H_d}+24g^2{\tilde m}^2_{H_d}
\bigg(\frac{2\Lambda}{M_c}\bigg) \nonumber \\ 
&-&12\frac{{\rm Tr}(Y_D Y_D^\dag Y_D Y_D^\dag)}{{\rm Tr}(Y_D Y_D^\dag)} 
\,{\tilde m}^2_{H_d} \bigg(\frac{2\Lambda}{M_c}\bigg)
\end{eqnarray}
where $a,b$ are $O(1)$ dimensionless quantities.
Thus, neglecting the first two log terms in the above equation
and using eq.~(\ref{rg1}),
we can find the approximate solution for the above equation as   
\begin{eqnarray}
\ln\bigg(\frac{{\tilde m}^2_{H_d} g^{\frac{16}{3}}(\Lambda)}
{{\tilde m}^2_{H_d} g^{\frac{16}{3}}(M_c)}\bigg) 
&\lesssim& \frac{4g^{-2}(M_c)}{3(2\eta_D-1)}
\frac{{\rm Tr}(Y_D Y_D^\dag Y_D Y_D^\dag)}{{\rm Tr}(Y_D Y_D^\dag)} 
\sim 10^{-4}
\end{eqnarray}
where we used the approximate solution for $Y_D$ in eq.~(7.20).
As a result, 
we can find the RG evolution of the down type Higgs mass negligible 
compared to other parameters with power running: 
${\tilde m}^2_{H_d} g^{\frac{16}{3}}$ is almost constant, so is 
$m^2_{H_d} \propto ({\rm Tr}Y_DY^\dag_D)\cdot g^{-\frac{16}{3}}\propto g^{4\eta_D-\frac{16}{3}}=g^0$.

By restoring the GUT breaking effect between the GUT scale and $M_c$
in the same way as in the previous section, 
we can find the following flavor conserving behavior of soft mass parameters 
in the infra-red limit and their deviations at $M_c$ as follows:
\begin{eqnarray}
\frac{B}{M\mu}(M_c)&=&-2(\eta_\mu-t_2\eta^X_\mu) \nonumber \\
&+&\bigg(\frac{g(M_{GUT})}{g(M_c)}\bigg)^2
\bigg(\frac{B}{M\mu}(M_{GUT})+2\eta_\mu\bigg), \\
\frac{h_U}{MY_U}(M_c)&=&-2(\eta_U-t_2\eta_U^X) \nonumber \\
&+&\bigg(\frac{g(M_{GUT})}{g(M_c)}\bigg)^2
\bigg(\frac{h_U}{MY_U}(M_{GUT})+2\eta_U\bigg)
+\delta_{h_U}(M_c), \\
\frac{h_D}{MY_D}(M_c)&=&-2(\eta_D-t_2\eta_D^X) \nonumber \\
&+&\bigg(\frac{g(M_{GUT})}{g(M_c)}\bigg)^2
\bigg(\frac{h_D}{MY_D}(M_{GUT})+2\eta_D\bigg)
+\delta_{h_D}(M_c), 
\end{eqnarray} 
and 
\begin{eqnarray}
\frac{(m^2_{10})^i_j}{|M|^2}(M_c)&=&\frac{4}{9}(C_2(10)-t_3C^X_2(10))\delta^i_j
\nonumber \\
&+&\bigg(\frac{g(M_{GUT})}{g(M_c)}\bigg)^4
\bigg(\frac{(m^2_{10})^i_j}{|M|^2}(M_{GUT})-\frac{4}{9}C_2(10)\delta^i_j\bigg)
+(\delta_{10})^i_j(M_c), \\
\frac{(m^2_{\bar 5})^i_j}{|M|^2}(M_c)&=&
\frac{4}{9}(C_2({\bar 5})-t_3C^X_2({\bar 5})) \delta^i_j \nonumber \\
&+&\bigg(\frac{g(M_{GUT})}{g(M_c)}\bigg)^4
\bigg(\frac{(m^2_{\bar 5})^i_j}{|M|^2}(M_{GUT})-\frac{4}{9}C_2({\bar 5})
\delta^i_j\bigg)+(\delta_{\bar 5})^i_j(M_c), \\
\frac{m^2_{H_u}}{|M|^2}(M_c)&=&
\frac{4}{9}(C_2(5)-t_3C^X_2(5)) \nonumber \\
&+&\bigg(\frac{g(M_{GUT})}{g(M_c)}\bigg)^4
\bigg(\frac{m^2_{H_u}}{|M|^2}(M_{GUT})-\frac{4}{9}C_2(5)\bigg),
\end{eqnarray}
where
\begin{eqnarray}
|\delta_{h_U}(M_c)|
&\simeq&\bigg|\frac{16}{9g^2(M_c)}\int^{g(M_{GUT})}_{g(M_c)}
\frac{g^2}{M}({\rm Tr}h_D Y_D^\dag)\frac{dg}{g^3}\bigg|, \nonumber \\
&\lesssim& \frac{8}{9}\frac{|{\rm Tr}Y_DY_D^\dag|}{g^2}(M_c) 
\sim 10^{-4}, \label{hcorr1}\\
|\delta_{h_D}(M_c)|
&\simeq&\bigg|\frac{8}{3g^2(M_c)}\int^{g(M_{GUT})}_{g(M_c)}
\frac{g^2}{M}({\rm Tr}h_D Y_D^\dag)\frac{dg}{g^3}\bigg|, \nonumber \\
&\lesssim& \frac{4}{3}\frac{|{\rm Tr}Y_DY_D^\dag|}{g^2}(M_c) 
\sim 10^{-4}\label{hcorr2},
\end{eqnarray}
and 
\begin{eqnarray} 
& &|(\delta_{10})^i_j(M_c)|
=\frac{1}{2}|(\delta_{\bar 5})^i_j(M_c)| \nonumber \\ 
&\simeq &\bigg|\frac{8}{9g^4(M_c)}\int^{g(M_{GUT})}_{g(M_c)}
\frac{g^4}{|M|^2}((h_Dh_D^\dag)^i_j
+(m^2_{10}+m^2_{\bar 5}+m^2_{H_d})(Y_D Y_D^\dag)^i_j)
\frac{dg}{g^3}\bigg| \nonumber \\
&\lesssim& \frac{|(Y_DY_D^\dag)^i_j|}{g^2}(M_c)\sim 10^{-4}.\label{mcorr} 
\end{eqnarray}
Here we find the different measure of the non-universal running 
of soft parameters compared to the case without bulk Higgs: 
$\eta^X_{ij}=1/3(2/9)$ for Higgs doublet(triplet),
$\eta^X_U=10/9$ and $\eta^X_D=7/9(1/3)$ for down type quark(lepton) Yukawa
coupling.
Note that when we made approximations for the corrections 
due to the bulk Higgs,
we used the approximate solutions for $Y_U$ and $Y_D$ given by eq.~(7.20) 
and the fact that 
$h_U\sim MY_U$ and $h_D\sim MY_D$ at $M_c$.
For instance, we inserted the dominant evolution of soft masses 
due to gauge interaction into the right-hand sides 
of eqs.~(\ref{hcorr1})-(\ref{mcorr}).
We also find that the dominant flavor violation 
due to the bulk Higgs comes from the KK modes near the compactification scale,
which can be seen from the scale-independent bounds on the corrections
as in eqs.~(\ref{hcorr1})-(\ref{mcorr}). The reason is that as the energy scale
increases, the Yukawa couplings powerly run into small values 
due to the dominant contribution from the KK modes of gauge fields, so the 
highest KK modes of the bulk Higgs are weakly coupled to the brane matters.
 
Consequently, we can find that independently of the precise separation 
between the GUT scale and the compactification scale, 
the additional corrections 
with flavor dependence to the soft masses of brane scalars 
are of order less than $10^{-4}$, which is small enough to
satisfy the experimental limits. When we assume that $h_U$($h_D$) is
of order $MY_U$($MY_D$) at the GUT scale, we also find that the generated
non-aligned components of trilinear couplings are of order less than $10^{-4}$ 
and $10^{-6}$ for $h_U$ and $h_D$, respectively, which are also small enough
for good agreement with experiments. 

On the other hand, we still get the suppression of 
phases due to the power corrections even with the bulk Higgs. 
In the case (1) which we are interested in, 
the phases of the trilinear couplings and the $B$ term are 
still suppressed to be of order $10^{-2}$ via 
the power running due to the bulk gauge multiplet. We also obtain 
the additional phases of trilinear couplings to be of order $10^{-4}$ 
due to the KK modes of the bulk Higgs multiplet. So, we end
up with the sufficiently suppressed $CP$ violation 
even in the case with one bulk Higgs.

\section{Conclusion}

We have shown that in 5D SUSY $SU(5)$ GUT on $S^1/Z_2$, the difference between
the GUT scale and the compactification scale gives 
the non-universal power running of soft masses such that their flavor 
dependent part can be negligible in the infrared limit. 
Firstly, for our purpose, 
we have taken the simple choice with only gauge fields in the bulk
and matter and Higgs fields on the brane. 
We included the GUT breaking effect for the renormalization group evolution of
soft masses 
and showed that it makes the difference of less than $O(1)$
in the fixed point values of soft mass parameters for the same GUT multiplet.
Next, we found that with the addition of
bulk Higgs field of down type, 
we not only explain 
the top-bottom mass hierarchy but also obtain a successful unification of 
gauge couplings due to the KK modes of the bulk Higgs multiplet. 
In either case without or with a bulk Higgs, 
the $O(1)$ $CP$ phases of soft mass parameters given at $M_{GUT}$ are  
sufficiently suppressed to be of order $10^{-2}$ 
to satisfy the stringent bounds from the EDMs.

\begin{figure}
\begin{center}
\begin{picture}(400,540)(-20,-40)
  \Line(10,0)(400,0)
  \LongArrow(10,0)(10,420)
  \Text(0,0)[r]{$0$}
  \Line(8,30)(12,30) \Text(6,60)[r]{$M_c$} \DashLine(10,30)(400,30){1}
  \Line(8,60)(12,60) \Text(6,120)[r]{$2M_c$}\DashLine(10,60)(400,60){1}
  \Line(8,90)(12,90) \Text(6,180)[r]{$3M_c$}\DashLine(10,90)(400,90){1}
  \Text(134,280)[b]{$0.2m_\Sigma$}
  \Text(325,280)[b]{$0.2m_\Sigma$}
  \Text(323,305)[b]{$m_\Sigma$}
  \Line(8,307)(12,307) \Text(6,307)[r]{$M_{GUT}$}

  \Line(8,120)(12,120)\DashLine(10,120)(400,120){1}
  \Line(8,150)(12,150)\DashLine(10,150)(400,150){1}
  \Line(8,180)(12,180)\DashLine(10,180)(400,180){1}
  \Line(8,210)(12,210)\DashLine(10,210)(400,210){1}
  \Line(8,240)(12,240)\DashLine(10,240)(400,240){1}
  \Line(8,270)(12,270)\DashLine(10,270)(400,270){1}
  \Line(8,300)(12,300)\DashLine(10,300)(400,300){1}
  \Line(8,330)(12,330)\DashLine(10,330)(400,330){1}
  \Line(8,360)(12,360)\DashLine(10,360)(400,360){1}
  \Line(8,390)(12,390)\DashLine(10,390)(400,390){1}
  \Text(6,420)[r]{$M_*$}

  \Text(60,-20)[b]{$(V^{SM},\lambda_1^{SM})$}

  \GBox(35,-1)(55,1){0}    \GBox(65,-1)(85,1){0}
  \GBox(35,59)(55,61){0}   \GBox(65,59)(85,61){0}
  \GBox(35,119)(55,121){0} \GBox(65,119)(85,121){0}
  \GBox(35,179)(55,181){0} \GBox(65,179)(85,181){0}
  \Vertex(45,199){1}\Vertex(75,199){1}
  \Vertex(45,209){1}\Vertex(75,209){1}
  \Vertex(45,219){1}\Vertex(75,219){1}

  \Text(120,-20)[b]{$(\lambda_2^{SM},\Phi^{SM})$}
  \GBox(95,59)(115,61){0}   \GBox(125,59)(145,61){0}
  \GBox(95,119)(115,121){0} \GBox(125,119)(145,121){0}
  \GBox(95,179)(115,181){0} \GBox(125,179)(145,181){0}
  \Vertex(105,199){1}\Vertex(135,199){1}
  \Vertex(105,209){1}\Vertex(135,209){1}
  \Vertex(105,219){1}\Vertex(135,219){1}

  \Text(180,-20)[b]{$(\tilde{\Sigma}^{SM},\Sigma^{SM})$}
  \GBox(155,281)(175,283){0} \GBox(185,281)(205,283){0}
\Text(120,-40)[b]{$\langle \Sigma \rangle =0$}
  \Text(235,-20)[b]{$V^{SM}$}
  \GBox(225,-1)(245,1){0}
  \GBox(225,59)(245,61){0}
  \GBox(225,119)(245,121){0}
  \GBox(225,179)(245,181){0}
  \Vertex(235,199){1}
  \Vertex(235,209){1}
  \Vertex(235,219){1}

  \Text(275,-20)[b]{$(\lambda_{1,2}^{SM})$}
  \GBox(265,4)(285,6){0}
  \GBox(265,64)(285,66){0}   \GBox(265,54)(285,56){0}
  \GBox(265,124)(285,126){0} \GBox(265,114)(285,116){0}
  \GBox(265,184)(285,186){0} \GBox(265,174)(285,176){0}
  \Vertex(275,199){1}
  \Vertex(275,209){1}
  \Vertex(275,219){1}

  \Text(310,-20)[b]{$\Phi^{SM}$}
  \GBox(300,59)(320,61){0}
  \GBox(300,119)(320,121){0}
  \GBox(300,179)(320,181){0}
  \Vertex(310,199){1}
  \Vertex(310,209){1}
  \Vertex(310,219){1}

  \Text(370,-20)[b]{$(\tilde{\Sigma}^{SM},\Sigma^{SM})$}
  \GBox(345,281)(365,283){0} \GBox(375,281)(395,283){0}
  \GBox(345,306)(365,308){0} \GBox(375,306)(395,308){0}

  \Text(310,-40)[b]{$\langle \Sigma \rangle \neq 0$}

\end{picture}
\caption{Mass spectrum of the SM group part before and after the GUT/SUSY
breaking on the brane.
After the GUT breaking, the $(8,1)$,$(1,3)$ Higgs and Higgsino has the
mass $m_\Sigma$ while the $(1,1)$ Higgs and Higgsino has the mass $0.2m_\Sigma$.
}
\end{center}
\end{figure}
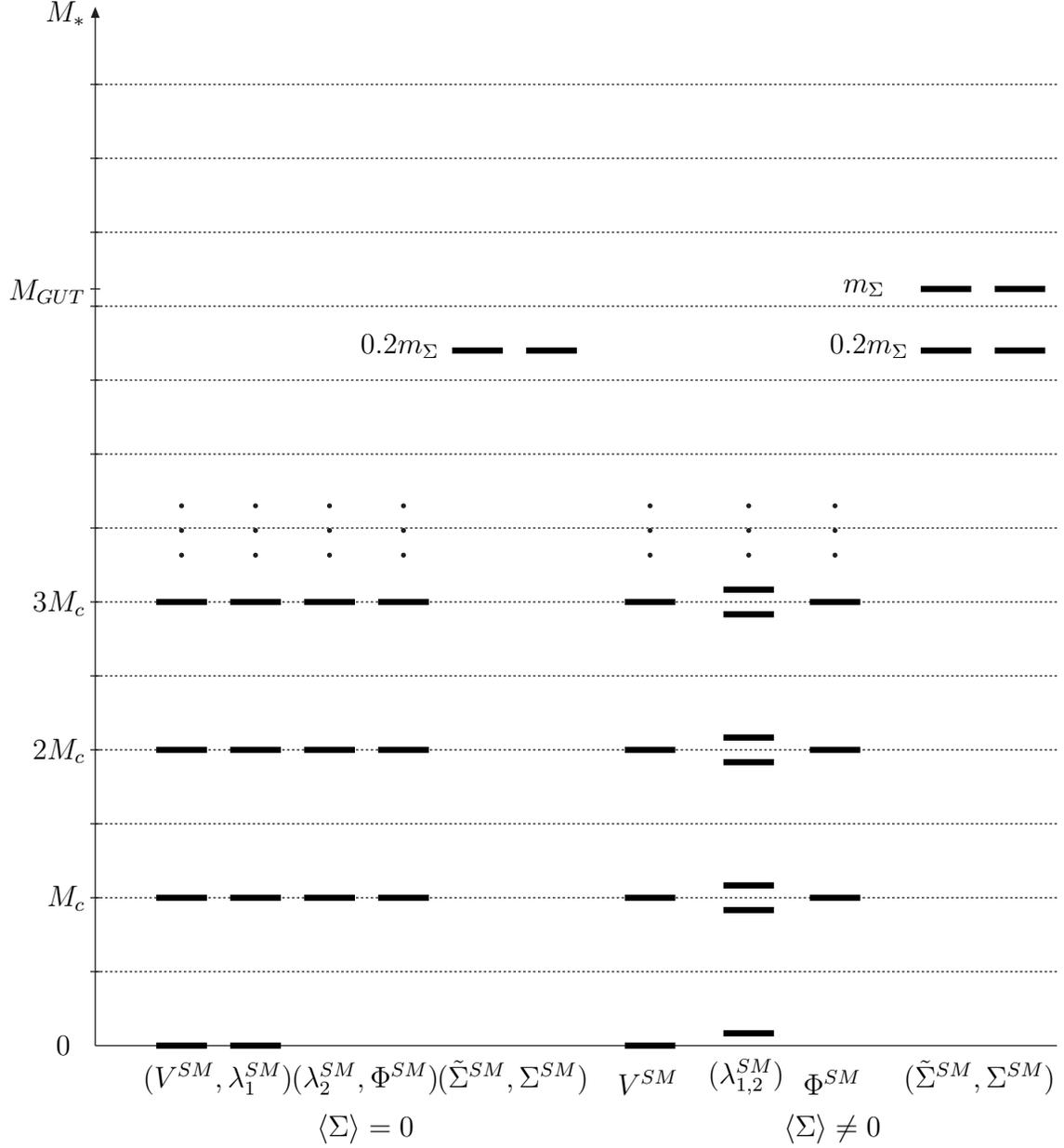

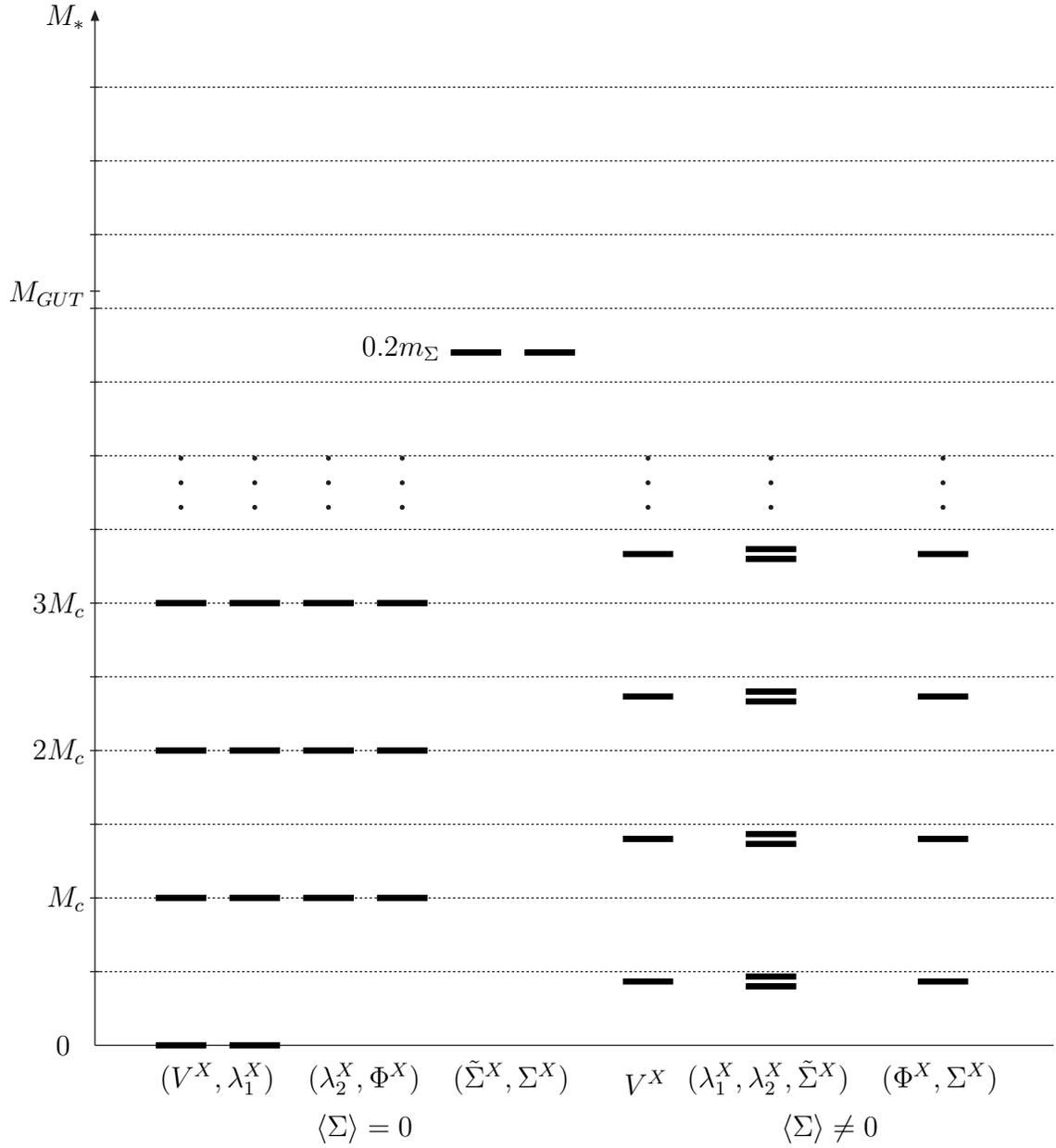
\begin{figure}
\begin{center}
\begin{picture}(400,540)(-20,-40)
  \Line(10,0)(400,0)
  \LongArrow(10,0)(10,420)
  \Text(0,0)[r]{$0$}
  \Line(8,30)(12,30) \Text(6,60)[r]{$M_c$} \DashLine(10,30)(400,30){1}
  \Line(8,60)(12,60) \Text(6,120)[r]{$2M_c$}\DashLine(10,60)(400,60){1}
  \Line(8,90)(12,90) \Text(6,180)[r]{$3M_c$}\DashLine(10,90)(400,90){1}
  \Line(8,307)(12,307) \Text(6,307)[r]{$M_{GUT}$}

  \Line(8,120)(12,120)\DashLine(10,120)(400,120){1}
  \Line(8,150)(12,150)\DashLine(10,150)(400,150){1}
  \Line(8,180)(12,180)\DashLine(10,180)(400,180){1}
  \Line(8,210)(12,210)\DashLine(10,210)(400,210){1}
  \Line(8,240)(12,240)\DashLine(10,240)(400,240){1}
  \Line(8,270)(12,270)\DashLine(10,270)(400,270){1}
  \Line(8,300)(12,300)\DashLine(10,300)(400,300){1}
  \Line(8,330)(12,330)\DashLine(10,330)(400,330){1}
  \Line(8,360)(12,360)\DashLine(10,360)(400,360){1}
  \Line(8,390)(12,390)\DashLine(10,390)(400,390){1} \Text(6,420)[r]{$M_*$}

  \Text(60,-20)[b]{$(V^{X},\lambda_1^{X})$}
  \GBox(35,-1)(55,1){0}    \GBox(65,-1)(85,1){0}
  \GBox(35,59)(55,61){0}   \GBox(65,59)(85,61){0}
  \GBox(35,119)(55,121){0} \GBox(65,119)(85,121){0}
  \GBox(35,179)(55,181){0} \GBox(65,179)(85,181){0}
  \Vertex(45,219){1}\Vertex(75,219){1}
  \Vertex(45,229){1}\Vertex(75,229){1}
  \Vertex(45,239){1}\Vertex(75,239){1}

  \Text(120,-20)[b]{$(\lambda_2^{X},\Phi^{X})$}
  \GBox(95,59)(115,61){0}   \GBox(125,59)(145,61){0}
  \GBox(95,119)(115,121){0} \GBox(125,119)(145,121){0}
  \GBox(95,179)(115,181){0} \GBox(125,179)(145,181){0}
  \Vertex(105,219){1}\Vertex(135,219){1}
  \Vertex(105,229){1}\Vertex(135,229){1}
  \Vertex(105,239){1}\Vertex(135,239){1}

  \Text(180,-20)[b]{$(\tilde{\Sigma}^{X},\Sigma^{X})$}
  \GBox(155,281)(175,283){0} \GBox(185,281)(205,283){0}

  \Text(120,-40)[b]{$\langle \Sigma \rangle =0$}
\Text(235,-20)[b]{$V^{X}$}
  \GBox(225,25)(245,27){0}
  \GBox(225,83)(245,85){0}
  \GBox(225,141)(245,143){0}
  \GBox(225,199)(245,201){0}
  \Vertex(235,219){1}
  \Vertex(235,229){1}
  \Vertex(235,239){1}

  \Text(285,-20)[b]{$(\lambda_1^{X},\lambda_2^{X},\tilde{\Sigma}^{X})$}
  \GBox(275,23)(295,25){0}   \GBox(275,27)(295,29){0}
  \GBox(275,81)(295,83){0}   \GBox(275,85)(295,87){0}
  \GBox(275,139)(295,141){0} \GBox(275,143)(295,145){0}
  \GBox(275,197)(295,199){0} \GBox(275,201)(295,203){0}
  \Vertex(285,219){1}
  \Vertex(285,229){1}
  \Vertex(285,239){1}

  \Text(355,-20)[b]{$(\Phi^{X},\Sigma^{X})$}
  \GBox(345,25)(365,27){0}
  \GBox(345,83)(365,85){0}
  \GBox(345,141)(365,143){0}
  \GBox(345,199)(365,201){0}
  \Text(135,280)[b]{$0.2m_\Sigma$}
  \Vertex(355,219){1}
  \Vertex(355,229){1}
  \Vertex(355,239){1}

  \Text(310,-40)[b]{$\langle \Sigma \rangle \neq 0$}

\end{picture}
\caption{Mass spectrum of the X,Y broken group part before
and after the GUT/SUSY breaking on the brane.}
\end{center}
\end{figure}

\begin{figure}
\begin{center}\begin{picture}(400,400)(0,0)
\PhotonArc(50,200)(30,0,180){2}{10}
\ArrowLine(-5,200)(20,200)
\ArrowLine(20,200)(80,200)
\ArrowLine(80,200)(105,200)
\Text(50,190)[]{$\phi$}
\Text(50,240)[]{$A$}
\Text(50,170)[]{$(a)$}
\PhotonArc(200,220)(20,0,360){2}{13}
\ArrowLine(150,200)(200,200)
\ArrowLine(200,200)(250,200)
\Text(200,190)[]{$\phi$}
\Text(200,250)[]{$A$}
\Text(200,170)[]{$(b)$}
\ArrowArc(350,200)(30,0,180)
\ArrowLine(295,200)(320,200)
\ArrowLine(320,200)(380,200)
\ArrowLine(380,200)(405,200)
\Text(300,190)[]{$\phi$}
\Text(350,190)[]{$\psi$}
\Text(400,190)[]{$\phi$}
\Text(350,240)[]{$\lambda$}
\Text(350,170)[]{$(c)$}
\DashCArc(105,80)(30,0,180){5}
\ArrowLine(50,80)(75,80)
\ArrowLine(75,80)(135,80)
\ArrowLine(135,80)(165,80)
\Text(155,70)[]{$\phi$}
\Text(65,70)[]{$\phi$}
\Text(105,70)[]{$\phi$}
\Text(105,120)[]{$\Phi$}
\Text(105,20)[]{$(d)$}
\ArrowArc(280,60)(20,100,80)
\ArrowLine(235,80)(275,80)
\ArrowLine(285,80)(325,80)
\Text(255,70)[]{$\phi$}
\Text(305,50)[]{$\phi$}
\Text(281,82)[]{$\centerdot$}
\Text(280,20)[]{$(e)$}
\end{picture}
\caption{One-loop Feynman diagrams 
for the self-energy of a brane scalar $\phi$.}
\end{center}
\label{fig:self}
\end{figure}
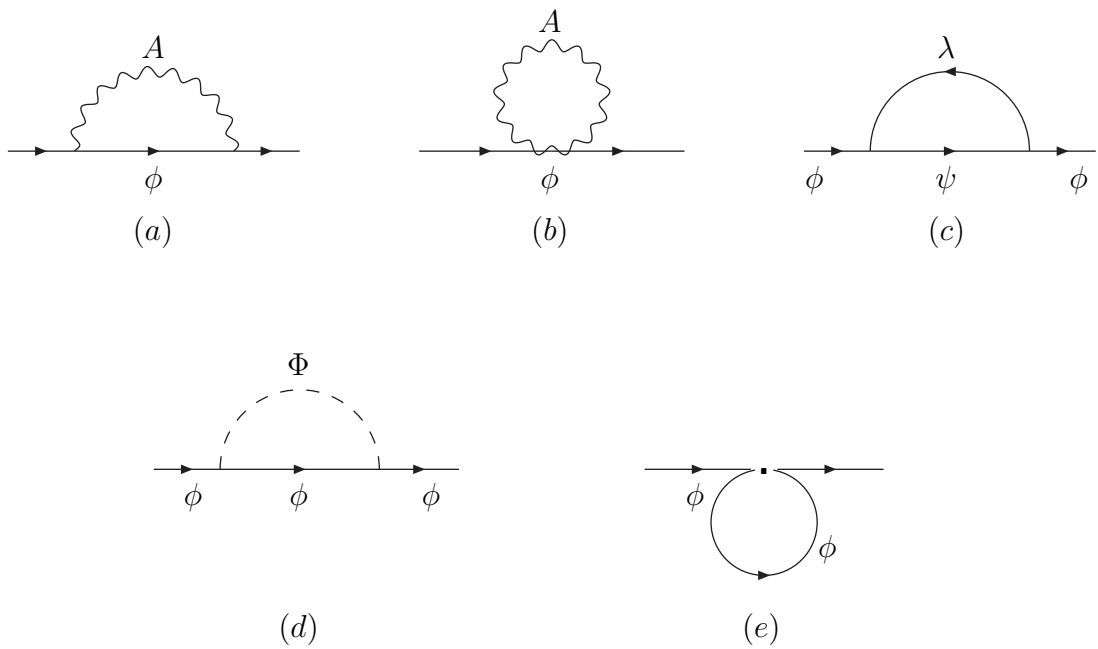

\bigskip

\acknowledgments
We would like to thank Haruhiko Terao for helpful comments on the manuscript. 
One of us(JEK) thanks Humboldt Foundation for the award. This work(KYC,JEK)
is supported in part by the BK21 program of Ministry of Education, KOSEF Sundo
Grant, and Korea Research Foundation Grant No. KRF-PBRG-2002-070-C00022.
This work(HML) is supported by the
European Community's Human Potential Programme under contracts
HPRN-CT-2000-00131 Quantum Spacetime, HPRN-CT-2000-00148 Physics Across the
Present Energy Frontier and HPRN-CT-2000-00152 Supersymmetry and the Early
Universe. HML was supported by priority grant 1096 of the Deutsche
Forschungsgemeinschaft.

\end{document}